\pdfoutput=1
\documentclass[12pt]{article}
\usepackage{graphics,graphicx}
\usepackage{amssymb,epsfig,amsmath,euscript,array}
\usepackage{cite}
\usepackage{pstricks}
\usepackage{color}

\makeatletter
\@addtoreset{equation}{section}
\makeatother

\newcounter{multieqs}

\newcommand{\be}{\begin{equation}}
\newcommand{\ee}{\end{equation}}

\newcommand{\bm}[1]{\mbox{\boldmath $#1$}}

\newcommand{\kslash}{k \!\!\! / }

\newcommand{\lslash}{l \!\! / }
\newcommand{\Pslash}{P \!\!\!\! / }

\newcommand{\islash}{i \!\!\! / }
\newcommand{\jslash}{j \!\!\! / }
\newcommand{\aslash}{a \!\!\! / }
\newcommand{\bslash}{{b \hspace{-6pt} \slash} }

\newcommand{\onslash}{1 \!\!\! / }
\newcommand{\twslash}{2 \!\!\!/ }
\newcommand{\thslash}{3 \!\!\!/ }
\newcommand{\foslash}{4 \!\!\! / }
\newcommand{\fislash}{5 \!\!\! / }

\newcommand{\mslash}{m \!\!\! / }

\def\bd{\begin{document}}
\def\ed{\end{document}}
\def\nn{\nonumber}
\def\bea{\begin{eqnarray}}
\def\eea{\end{eqnarray}}

\def\red{\color{red}}
\def\black{\color{black}}
\def\blue{\color{blue}}
\def\orange{\color{orange}}

\def\ab{(ijab)}
\def\ba{(ijba)}
\def\ijab{{\tr}_{+}(\islash\, \jslash\, \aslash \, \bslash)}
\def\ijba{{\tr}_{+}(\islash\, \jslash\, \bslash \, \aslash)}
\def\ijaP{{\tr}_{+}(\islash\, \jslash\, \aslash \, \Pslash)}
\def\ijPLa{{\tr}_{+}(\islash\, \jslash\, \Pslash_L \, \aslash)}
\def\ijaPL{{\tr}_{+}(\islash\, \jslash\, \aslash \, \Pslash_L)}
\def\ijPLza{{\tr}_{+}(\islash\, \jslash\, \Pslash_{L;z} \, \aslash)}
\def\ijaPLz{{\tr}_{+}(\islash\, \jslash\, \aslash \, \Pslash_{L;z})}
\def\ijPa{{\tr}_{+}(\islash\, \jslash\, \Pslash \, \aslash)}
\def\iaPb{{\tr}_{+}(\islash\, \aslash\, \Pslash \, \bslash)}
\def\ibPa{{\tr}_{+}(\islash\, \bslash\, \Pslash \, \aslash)}
\def\ijPmu{{\tr}_{+}(\islash\, \jslash\, \Pslash \, \mu)}
\def\ibmuP{{\tr}_{+}(\islash\, \bslash\, \mu \, \Pslash)}
\def\ibmua{{\tr}_{+}(\islash\, \bslash\, \mu \, \aslash)}
\def\iamub{{\tr}_{+}(\islash\, \aslash\, \mu \, \bslash)}
\def\jaPb{{\tr}_{+}(\jslash\, \aslash\, \Pslash \, \bslash)}
\def\ijmuP{{\tr}_{+}(\islash\, \jslash\, \mu \, \Pslash)}
\def\ijmum{{\tr}_{+}(\islash\, \jslash\, \mu \, \mslash)}
\def\ijmmu{{\tr}_{+}(\islash\, \jslash\, \mslash \, \mu)}
\def\ijmP{{\tr}_{+}(\islash\, \jslash\, \mslash \, \Pslash)}
\def\iabP{{\tr}_{+}(\islash\, \aslash\, \bslash \, \Pslash)}
\def\ijbP{{\tr}_{+}(\islash\, \jslash\, \bslash \, \Pslash)}
\def\jbPa{{\tr}_{+}(\jslash\, \bslash\, \Pslash \, \aslash)}
\def\ijPb{{\tr}_{+}(\islash\, \jslash\, \Pslash \, \bslash)}
\def\jbmua{{\tr}_{+}(\jslash\, \bslash\, \mu \, \aslash)}

\def\loablt{ {\tr}_{+}(\lslash_1\, \aslash \, \bslash\, \lslash_2)}

\def\ijlolt{{\tr}_{+}(\islash\, \jslash\, \lslash_1 \, \lslash_2)}
\def\ijltlo{{\tr}_{+}(\islash\, \jslash\, \lslash_2 \, \lslash_1)}
\def\ibloa{{\tr}_{+}(\islash\, \bslash\, \lslash_1 \, \aslash)}
\def\jaltb{{\tr}_{+}(\jslash\, \aslash\, \lslash_2 \, \bslash)}
\def\ialtb{{\tr}_{+}(\islash\, \aslash\, \lslash_2 \, \bslash)}
\def\bltloa{{\tr}_{+}(\bslash\, \lslash_2\, \lslash_1 \, \aslash)}
\def\jbloa{{\tr}_{+}(\jslash\, \bslash\, \lslash_1 \, \aslash)}
\def\ibPb{{\tr}_{+}(\islash\, \bslash\, \Pslash \, \bslash)}
\def\ijltb{{\tr}_{+}(\islash\, \jslash\, \lslash_2 \, \bslash)}

\def\ijloa{{\tr}_{+}(\islash\, \jslash\,  \lslash_1 \, \aslash)}
\def\ijblt{{\tr}_{+}(\islash\, \jslash\,  \bslash \, \lslash_2)}

\def\jakb{{\tr}_{+}(\jslash\, \aslash\, \kslash \, \bslash)}
\def\iakb{{\tr}_{+}(\islash\, \aslash\, \kslash \, \bslash)}

\def\tofo{{\tr}_{+}(\onslash\, \thslash\, \twslash \, \foslash)}
\def\foto{{\tr}_{+}(\onslash\, \thslash\, \foslash \, \twslash)}
\def\tofi{{\tr}_{+}(\onslash\, \thslash\, \twslash \, \fislash)}
\def\fito{{\tr}_{+}(\onslash\, \thslash\, \fislash \, \twslash)}

\def\lrangle#1#2{\langle #1\,#2\rangle}

\def\Li{{$\rm Li}_2$}
\def\eps{\epsilon}
\def\epsuv{{\epsilon_{\rm \mbox{\tiny UV}}}}
\let\bm=\bibitem
\let\la=\label

\def\npb#1#2#3{Nucl. Phys. {\bf{B#1}} #3 (#2)}
\def\plb#1#2#3{Phys. Lett. {\bf{#1B}} #3 (#2)}
\def\prl#1#2#3{Phys. Rev. Lett. {\bf{#1}} #3 (#2)}
\def\prd#1#2#3{Phys. Rev. {D \bf{#1}} #3 (#2)}
\def\cmp#1#2#3{Comm. Math. Phys. {\bf{#1}} #3 (#2)}
\def\cqg#1#2#3{Class. Quantum Grav. {\bf{#1}} #3 (#2)}
\def\nppsa#1#2#3{Nucl. Phys. B (Proc. Suppl.) {\bf{#1A}}#3 (#2)}
\def\ap#1#2#3{Ann. of Phys. {\bf{#1}} #3 (#2)}
\def\ijmp#1#2#3{Int. J. Mod. Phys. {\bf{A#1}} #3 (#2)}
\def\rmp#1#2#3{Rev. Mod. Phys. {\bf{#1}} #3 (#2)}
\def\mpla#1#2#3{Mod. Phys. Lett. {\bf A#1} #3 (#2)}
\def\jhep#1#2#3{J. High Energy Phys. {\bf #1} #3 (#2)}
\def\atmp#1#2#3{Adv. Theor. Math. Phys. {\bf #1} #3 (#2)}
\newcommand{\EQ}[1]{\begin{equation} #1 \end{equation}}
\newcommand{\AL}[1]{\begin{subequations}\begin{align} #1 \end{align}\end{subequations}}
\newcommand{\SP}[1]{\begin{equation}\begin{split} #1 \end{split}\end{equation}}
\newcommand{\ALAT}[2]{\begin{subequations}\begin{alignat}{#1} #2 \end{alignat}
                        \end{subequations}}
\def\beqa{\begin{eqnarray}}
\def\eeqa{\end{eqnarray}}
\def\beq{\begin{equation}}
\def\eeq{\end{equation}}
\def\sst{\scriptscriptstyle}
\def\thetabar{\bar\theta}
\def\Tr{{\rm Tr}}
\def\one{\mbox{1 \kern-.59em {\rm l}}}
 \def\Nh{\hat{N}}

\newcommand{\half}{{\textstyle {1 \over 2}}}

\def\a{\alpha}      \def\da{{\dot\alpha}}
\def\b{\beta}       \def\db{{\dot\beta}}
\def\c{\gamma}  \def\G{\Gamma}  \def\cdt{\dot\gamma}
\def\d{\delta}  \def\D{\Delta}  \def\ddt{\dot\delta}
\def\e{\epsilon}        \def\vare{\varepsilon}
\def\f{\phi}    \def\F{\Phi}    \def\vvf{\f}
\def\h{\eta}
\def\k{\kappa}
\def\l{\lambda} \def\L{\Lambda}
\def\m{\mu} \def\n{\nu}
\def\o{\omega}
\def\p{\pi} \def\P{\Pi}
\def\r{\rho}
\def\s{\sigma}  \def\S{\Sigma}
\def\t{\tau}
\def\th{\theta} \def\Th{\Theta} \def\vth{\vartheta}
\def\X{\Xeta}
\def\z{\zeta}
\def\de{\partial}

\def\cA{{\cal A}} \def\cB{{\cal B}} \def\cC{{\cal C}}
\def\cD{{\cal D}} \def\cE{{\cal E}} \def\cF{{\cal F}}
\def\cG{{\cal G}} \def\cH{{\cal H}} \def\cI{{\cal I}}
\def\cJ{{\cal J}} \def\cK{{\cal K}} \def\cL{{\cal L}}
\def\cM{{\cal M}} \def\cN{{\cal N}} \def\cO{{\cal O}}
\def\cP{{\cal P}} \def\cQ{{\cal Q}} \def\cR{{\cal R}}
\def\cS{{\cal S}} \def\cT{{\cal T}} \def\cU{{\cal U}}
\def\cV{{\cal V}} \def\cW{{\cal W}} \def\cX{{\cal X}}
\def\cY{{\cal Y}} \def\cZ{{\cal Z}}

\def\ua{\underline{\alpha}}
\def\ub{\underline{\phantom{\alpha}}\!\!\!\beta}
\def\uc{\underline{\phantom{\alpha}}\!\!\!\gamma}
\def\um{\underline{\mu}}
\def\ud{\underline\delta}
\def\ue{\underline\epsilon}
\def\una{\underline a}\def\unA{\underline A}
\def\unb{\underline b}\def\unB{\underline B}
\def\unc{\underline c}\def\unC{\underline C}
\def\und{\underline d}\def\unD{\underline D}
\def\une{\underline e}\def\unE{\underline E}
\def\unf{\underline{\phantom{e}}\!\!\!\! f}\def\unF{\underline F}
\def\unm{\underline m}\def\unM{\underline M}
\def\unn{\underline n}\def\unN{\underline N}
\def\unp{\underline{\phantom{a}}\!\!\! p}\def\unP{\underline P}
\def\unq{\underline{\phantom{a}}\!\!\! q}
\def\unQ{\underline{\phantom{A}}\!\!\!\! Q}
\def\unH{\underline{H}}

\def\As {{A \hspace{-6.4pt} \slash}\;}
\def\bs {{b \hspace{-6.4pt} \slash}\;}
\def\Ds {{D \hspace{-6.4pt} \slash}\;}
\def\ds {{\del \hspace{-6.4pt} \slash}\;}
\def\ss {{\s \hspace{-6.4pt} \slash}\;}
\def\ks {{ k \hspace{-6.4pt} \slash}\;}
\def\ps {{p \hspace{-6.4pt} \slash}\;}
\def\pas {{{p_1} \hspace{-6.4pt} \slash}\;}
\def\pbs {{{p_2} \hspace{-6.4pt} \slash}\;}
\def\Ps {{P \hspace{-6.4pt} \slash}\;}
\def\Qs {{Q \hspace{-6.4pt} \slash}\;}

\def\Fh{\hat{F}}
\def\Vh{\hat{V}}
\def\Xh{\hat{X}}
\def\ah{\hat{a}}
\def\xh{\hat{x}}
\def\yh{\hat{y}}
\def\ph{\hat{p}}
\def\xih{\hat{\xi}}
\def\psit{\tilde{\psi}}
\def\Psit{\tilde{\Psi}}
\def\tht{\tilde{\th}}
\def\lt{\tilde{\lambda}}
\def\hl{\hat{\lambda}}
\def\hlt{\hat{\tilde{\lambda}}}
\def\llt{\tilde{l}}
\def\At{\tilde{A}}
\def\Qt{\tilde{Q}}
\def\Rt{\tilde{R}}
\def\Nt{\tilde{N}}

\def\at{\tilde{a}}
\def\st{\tilde{s}}
\def\ft{\tilde{f}}
\def\pt{\tilde{p}}
\def\qt{\tilde{q}}
\def\vt{\tilde{v}}
\def\nt{\tilde{n}}

\def\delb{\bar{\partial}}
\def\bz{\bar{z}}
\def\bD{\bar{D}}
\def\bB{\bar{B}}

\def\bk{{\bf k}}
\def\bl{{\bf l}}
\def\bp{{\bf p}}
\def\bq{{\bf q}}
\def\br{{\bf r}}
\def\bx{{\bf x}}
\def\by{{\bf y}}
\def\bR{{\bf R}}
\def\bV{{\bf V}}

\def\d{\delta}\def\D{\Delta}\def\ddt{\dot\delta}
\def\pa{\partial} \def\del{\partial}
\def\xx{\times}
\def\uno{\mbox{1 \kern-.59em {\rm l}}}
\def\trp{^{\top}}
\def\inv{^{-1}}
\def\dag{{^{\dagger}}}
\def\pr{^{\prime}}
\def\lan{\langle}
\def\ran{\rangle}
\def\rar{\rightarrow}
\def\lar{\leftarrow}
\def\lrar{\leftrightarrow}
\newcommand{\0}{\,\!}      
\def\one{1\!\!1\,\,}
\def\im{\imath}
\def\jm{\jmath}
\newcommand{\tr}{\mbox{tr}}
\newcommand{\slsh}[1]{/ \!\!\!\! #1}
\def\vac{|0\rangle}
\def\lvac{\langle 0|}
\def\hlf{\frac{1}{2}}
\def\ove#1{\frac{1}{#1}}
\def\Box{\square}
\def\ZZ{\mathbb{Z}}
\def\CC#1{({\bf #1})}
\def\bcomment#1{}
\def\bfhat#1{{\bf \hat{#1}}}
\def\VEV#1{\left\langle #1\right\rangle}
\newcommand{\ex}[1]{{\rm e}^{#1}} \def\ii{{\rm i}}
\def\rr{{\rm r}} \def\rs{{\rm s}}\def\rv{{\rm v}}
\def\ri{{\rm i}}\def\rj{{\rm j}}
\newcommand{\lrbrk}[1]{\left(#1\right)}
\newcommand{\sfrac}[2]{{\textstyle\frac{#1}{#2}}}

\def\Li{{\rm Li}_2}

\font\mybb=msbm10 at 12pt
\def\bb#1{\hbox{\mybb#1}}

\font\myBB=msbm10 at 18pt
\def\BB#1{\hbox{\myBB#1}}

\begin{document}

\begin{flushright}
IPPP/09/83,
DCPT/09/166,
QMUL-PH-09-23
\end{flushright}

\vspace{3pt}

\begin{center}

\hspace{-0.8cm}{\Large \bf Simplicity of Polygon Wilson Loops in $\mathcal{N}=4$ SYM}

\vspace{11pt}
\end{center}

\hspace{-1.4cm}{\mbox {\bf   Andreas~Brandhuber$^{a}$, Paul~Heslop$^{b}$,
Valentin~V.~Khoze$^{c}$ and Gabriele~Travaglini$^{a}$}}%
\footnote{
{\tt  \{ \tt \!\!\!a.brandhuber,  g.travaglini\}@qmul.ac.uk, \{paul.heslop, valya.khoze\}@durham.ac.uk}
}

\begin{center}
{\small \em
\begin{itemize}
\item[\ \ \ \ \ \ $^a$]
Centre for Research in String Theory\\
Department of Physics,
Queen Mary University of London\\
Mile End Road, London, E1 4NS,
United Kingdom\\
\item[\ \ \ \ \ \ $^b$]
Institute for Particle Physics Phenomenology,  \\
Department of Mathematical Sciences and Department of Physics\\
Durham University,
Durham, DH1 3LE, United Kingdom\\
\item[\ \ \ \ \ \ $^c$]
Institute for Particle Physics Phenomenology,  \\
Department of Physics,
Durham University, \\
Durham, DH1 3LE, United Kingdom

\end{itemize}
}

\vspace{-8pt}

\vspace{23pt} {\bf Abstract}
\end{center}

\noindent
Wilson loops with lightlike polygonal contours have been conjectured to be equivalent to MHV scattering amplitudes
in ${\cal N}=4$ super Yang-Mills. We compute such Wilson loops for special polygonal contours at two loops in perturbation theory.
Specifically, we concentrate on the remainder function $\mathcal{R}$, obtained by subtracting the known ABDK/BDS ansatz from the Wilson loop.
First, we consider a particular two-dimensional eight-point kinematics studied at strong coupling by Alday and Maldacena.
We find numerical evidence that $\mathcal{R}$ is the same at weak and at strong coupling, up to an overall,
coupling-dependent constant.
This suggests a universality of the remainder function at strong and weak coupling for generic null polygonal Wilson loops,
and therefore for arbitrary MHV amplitudes in $\cN=4$ super Yang-Mills. We analyse the consequences of this statement.
We further consider regular $n$-gons, and find that the remainder function
is  linear in  $n$  at large $n$ through numerical computations performed up to $n=30$.
This reproduces a general feature of the corresponding strong-coupling result.

\setcounter{page}{0}
\thispagestyle{empty}
\newpage


\section{Introduction }
\setcounter{footnote}{0}

Scattering amplitudes in gauge theories, and in particular in $\cN=4$ super Yang-Mills (SYM), exhibit many
beautiful properties which have greatly expanded our understanding of their mathematical structure.
In \cite{abdk}, a remarkable relation was discovered, expressing the planar two-loop four-point amplitude in $\cN=4$ SYM
in terms of the one-loop amplitude. The authors of \cite{abdk} also conjectured a generalisation of this relation for two-loop
MHV amplitudes with an arbitrary number of external legs.
The possibility of iterative relations at higher loop order was investigated in \cite{bds}, and confirmed for three-loop, four-point
amplitudes. Furthermore, an intriguing conjecture was put forward in \cite{bds}, expressing an  appropriately defined finite part
of the planar all-loop $n$-point MHV amplitude as the exponential of the one-loop amplitude times the cusp anomalous dimension
\cite{Polyakov:1980ca,Brandt:1981kf,Korchemsky:1985xj}. An equation for the all-loop cusp anomalous dimension was found in \cite{bes},
and the strong-coupling expansion of its solution presented in \cite{bkk}.

In a remarkable development, Alday and Maldacena \cite{am} were able to calculate the four-point amplitude in $\cN=4$ super Yang-Mills at strong coupling using the AdS/CFT correspondence. Their result confirmed the structure of the ABDK/BDS proposal at four points,
with a simple replacement of the weak-coupling expression of the cusp anomalous dimension with its strong-coupling counterpart.
The prescription of \cite{am} maps the calculation of scattering amplitudes to that of a Wilson loop whose contour is obtained by gluing the particle momenta following  the ordering of the particles in the corresponding colour-stripped, planar amplitude. This suggested a remarkable duality, according to which MHV amplitudes are equal to expectation values of lightlike Wilson loops \cite{am, dks, bht}. This duality was tested at one loop for four particles in \cite{dks}, and for any number of particles in \cite{bht}. Further calculations performed in \cite{dhks4, dhks5} confirmed the duality at two loops in the four- and five-point case.
The iteration of the parity-even part of the five-point amplitude was checked in \cite{Cachazo:2006tj}.
This result was extended to the complete amplitude in  \cite{2l5pt}, where a
cancellation  of contributions arising from parity-odd terms in the logarithm of the amplitude was discovered.

The BDS ansatz was further studied in  \cite{Alday:2007he}, where it was found  that,
for a particular kinematic configuration with a large number of scattered particles, a certain discrepancy function -- now known
as the remainder function --  should be added to the BDS ansatz in order to reproduce the complete amplitude.
The same conclusion was also reached for the six-point case, based on perturbative Wilson loop calculations in \cite{dhksbum}
and the study of Regge limits in \cite{Bartels:2008ce}.
The remainder function was then obtained  by  an explicit two-loop calculation
of the parity-even part of the six-point MHV amplitude \cite{seven}. In \cite{dhksbum,dhks6} the corresponding hexagon
Wilson loop was computed, and the comparison of the amplitude and Wilson loop remainder functions
performed in \cite{seven,dhks6} showed that these two quantities were identical in all kinematical configurations considered.

One key goal is to characterise and eventually determine this remainder function.
An important insight of \cite{dhks5} was that the Wilson loop remainder function must be invariant under dual conformal symmetry. At four and five points, the lightlike condition on the particle momenta do not leave room for non-trivial conformally invariant cross-ratios, and the BDS conjecture remains intact.
Only starting from six points one can write down such cross-ratios, and dual conformal symmetry implies that the remainder function depends on the kinematic variables only through these cross-ratios \cite{dhks5}.

Additional information about this function was found in
\cite{seven,Anastasiou:2009kn} using simple and multi-collinear limits of amplitudes and Wilson loops. Furthermore,
Anastasiou, Spence and the present authors developed  in \cite{Anastasiou:2009kn} a numerical algorithm to evaluate
Wilson loop remainder functions with an arbitrary number of edges, based on  programs developed
in \cite{Anastasiou:2008rm, Anastasiou:2007qb, Lazopoulos:2007ix, Anastasiou:2005pn}
for evaluating Feynman diagrams. The results of \cite{Anastasiou:2009kn} make it possible to compare the Wilson loop remainder function to the  amplitude remainder function, once the integrals of the latter are known to the required order in the Laurent expansion in the dimensional regularisation parameter $\e$. We note that recently the parity-even part of an arbitrary two-loop $n$-point MHV amplitude has been written down in \cite{cristian-n} in terms of integral functions.
Particular limits of the two-loop Wilson loop remainder function were also recently studied in
\cite{Georgiou:2009mp, Schabinger:2009bb}.

In a very interesting recent development   \cite{amsmalloctopus,amoctopus},  a new class of Wilson loops was computed at strong coupling using results of \cite{Gaiotto:2008cd}.
In the AdS/CFT set-up of \cite{am}, the problem of calculating the amplitude at strong coupling is equivalent to that of minimising the area of a string worldsheet  bounded by a lightlike, polygonal contour living on the boundary of the ${\rm AdS}_5$ space.
In \cite{amoctopus}, Alday and Maldacena considered a more tractable situation where the contour of the Wilson loop is embedded in a  $\mathbb{R}^{1,1}$ subspace of the boundary of ${\rm AdS}_5$, with the worldsheet living in an ${\rm AdS}_3$ subspace of ${\rm AdS}_5$. The simplification achieved in this way is quite dramatic -- the number of independent
cross-ratios for a $2n$-gon contour is reduced to $2n-6$, where dual conformal symmetry has been used to map the position of three cusps  at infinity. This eliminates six out of the $2n$ parameters needed to fix the position of the $2n$ cusps.
Hence, at six points there are no free cross-ratios and the remainder function is a constant. Starting from $n=4$, or eight points, we  have non-trivial octagon kinematics determined by  two cross-ratios. In \cite{amoctopus}, the explicit form of the octagon remainder as a function of these two cross-ratios was presented in terms of a simple one-dimensional integral (up to  constant terms).

The original motivation of this paper was to evaluate at two loops the octagon remainder function in the kinematics of \cite{amoctopus}
using the techniques  of \cite{Anastasiou:2009kn} and to compare with the strong coupling expression found in \cite{amoctopus}. It is far from obvious that there should exist any simple relation between the strong and weak coupling limits of the remainder function.
However,
the results we find suggest that, in fact, the remainder functions at weak and strong coupling are linked by a simple linear relation,
\beq
\label{uno}
\cR_{8}^{ (2)} (u_{ij})\, =\, A^{(2)}   \, \cR_{8}^{\, \rm AM}
(u_{ij})\, +\, {B}^{(2)}  \ ,
\eeq
where $\cR_{8}^{ (2)} (u_{ij})$ is the two-loop remainder function,
and $ \cR_{8}^{\, \rm AM}$ is the strong-coupling remainder evaluated in \cite{amoctopus}; the cross-ratios are denoted by $u_{ij}$.
Using simple- and multi-collinear limits, we will further argue that the constant ${B}^{(2)}$
can be absorbed into a redefinition of the strong-coupling remainder function, which in practice leaves only one coupling-dependent
quantity, $A^{(2)}$, in the relation linking strong and weak coupling. We should stress that our calculation is based on numerical
evaluations, and is therefore correct only within errors.

If the correspondence does hold, it suggests the following general matching relation between the all-orders perturbative
remainder function, $\cR (a)$, and the strong-coupling expression, $\cR_{\rm strong}$,
\beq
\label{boldconjecture2}
\cR (a)  \, = \, A (a) \cR_{\rm strong}
\ ,
\eeq
where $a$ is the 't Hooft coupling, and the function $A (a) $ has a perturbative expansion starting  at two loops and is independent
of $n$ and of the kinematic variables
(which are all contained in $\cR_{\rm strong}$).
If \eqref{boldconjecture2} is correct, it should be very interesting to
find the physical meaning of the proportionality constant~$A(a)$.
Since the BDS part of the amplitude already has a universal structure at weak and strong coupling, combined
with \eqref{boldconjecture2}, this implies the universal structure of the Wilson loop (MHV amplitude) itself.

The second special class of kinematic configuration addressed by \cite{amoctopus} is that of regular $2n$-gons.
In this case the remainder function is just a number, and one can study it as a function of $n$.
Our main result here is that, at large $n$,  the regular remainder function at two loops is linear in $n$.
In this case we do not perform a full comparison of our weak-coupling results to the strong-coupling results
of \cite{amoctopus}, where only one part of the complete remainder function was presented.
It would clearly be of great interest to check whether a relation such as \eqref{boldconjecture2} still holds.
This could lend further support to the conjecture
that \eqref{boldconjecture2} is valid not only for the special octagon kinematics
embedded in $\mathbb{R}^{1,1}$, but for all $n$-gon Wilson loops, and for the corresponding $n$-point MHV scattering amplitudes in
$\cN=4$ SYM. If this is correct, the remainder function  is universal.

The rest of the paper is organised as follows. In Section 2, we review some basic facts about the Wilson loop remainder function,
in particular its precise definition and its behaviour in (multi-)collinear limits. Section 3 is devoted to the calculation of the
octagon remainder function at two loops. We begin by first reviewing the (1+1)-dimensional kinematics introduced in \cite{amoctopus},
and map it to conformally equivalent (2+1)-dimensional configurations where all the positions of the cusps are at finite distance.
Thanks to dual conformal invariance of the remainder function \cite{dhks5}, all these configurations are equivalent for the sake of
calculating this quantity, however our (2+1)-configurations are more convenient for numerical simulations.
We then compare the results at strong and weak coupling, presenting evidence in favour of \eqref{uno}, as well as estimates
of the parameters $A^{(2)}$ and $B^{(2)}$ appearing in that relation. In Section 4 we analyse the consequences of the
universality conjecture for the remainder function, and discuss how to reabsorb the constant $B^{(2)}$ into a redefinition
of the strong-coupling remainder function. Section 5 contains our results for regular $2n$-gons. Finally, we present
our conclusions in Section 6.

\section{Definition of the remainder function}
\label{review}

We expand the Wilson loop in powers of the 't Hooft coupling $a$ as%
\footnote{We follow the definitions and conventions of \cite{Anastasiou:2009kn}, to which we refer the reader for more details.}
\beq
\label{nae}
\lan W[\cC_n ] \ran \  :=  1 \, + \, \sum_{l=1}^{\infty} a^l W^{(l)}_n \ := \   \exp \sum_{l=1}^{\infty} a^l w^{(l)}_n
\ .
\eeq
In particular,
\beq
w^{(1)}_n \ = \  W^{(1)}_n \, , \qquad
w^{(2)}_n \ = \  W^{(2)}_n \, - \, {1\over2} \, (W^{(1)}_n)^2
\ .
\eeq
We note that $w^{(1)}_n$ times the tree-level MHV amplitude is equal to the one-loop MHV amplitude first calculated in \cite{bddk} using the unitarity-based approach \cite{fusing}, up to a regularisation-dependent factor. $w^{(1)}_n$ was calculated in \cite{dks} for $n=4$ and in \cite{bht} for arbitrary  $n$, while $w^{(2)}_n$ was evaluated analytically for $n=4$  in \cite{dhks4}, and numerically for $n=5, 6$ in \cite{dhks5,dhksbum,dhks6,Anastasiou:2009kn}. Results for $n=7,8$ were also presented in \cite{Anastasiou:2009kn}.

We define the remainder function $\cR_n^{\rm WL}$ for an $n$-sided Wilson loop  via
\begin{align}
\label{wbds}
  w_n^{(2)}(\e)  &= f_\mathrm{WL}^{(2)}(\epsilon) \, w_n^{(1)}(2 \epsilon) \, + \, C_\mathrm{WL}^{(2)} \, +\, \cR_n^\mathrm{WL}
  \  ,
\end{align}
where $\eps<0$, and
\beq \label{flepsW}
f_\mathrm{WL}^{(2)}(\epsilon ) :=\, f_0^{(2)} + f_{1,\mathrm{WL}}^{(2)} \epsilon + f_{2,\mathrm{WL}}^{(2)} \epsilon^2
\ .
\eeq
 $f_0^{(2)}$  is the same as on the amplitude side,  $f_0^{(2)}= \zeta_2$,
while $f_{1,\mathrm{WL}}^{(2)}= G_{\rm eik}^{(2)} = 7 \zeta_3 $  \cite{kk}.
In \cite{Anastasiou:2009kn}, the four- and five-edged Wilson loops were cast in the form \eqref{wbds}
with
\beq
\label{R45}
\cR_4^\mathrm{WL}\, = \, \cR_5^\mathrm{WL} \, = \, 0
\ ,
\eeq
which allowed for a determination of the coefficients $f_{2,\mathrm{WL}}^{(2)}$ and $C_\mathrm{WL}^{(2)}$.
The results found in\cite{Anastasiou:2009kn}, are%
\footnote{The $\cO (1)$ and $\cO (\e )$ coefficients of $f_\mathrm{WL}^{(2)}(\epsilon)$ had been determined earlier in
\cite{dhks4}.}
\beq
f_\mathrm{WL}^{(2)}(\epsilon)\ =\  -\zeta_2 \, + \,   7 \zeta_3 \, \e  \ - \    5\zeta_4\, \e^2
\ ,
\eeq
and
\beq
C_\mathrm{WL}^{(2)} \ =\  -{1\over 2} \zeta_2^2
\ .
\eeq
As noticed in  \cite{Anastasiou:2009kn}, there is an intriguing agreement between
the constant $C_\mathrm{WL}^{(2)}$ and the corresponding value
of the same quantity on the amplitude side.

The statement of the duality between Wilson loops and amplitudes  beyond one loop can be recast
as an  equality of the corresponding remainder functions,
\begin{align}
\label{dualitywa}
 \cR_n \ = \   \cR_n^\mathrm{WL}\ .
\end{align}
A consequence of the precise  determination of the  constants  $ f_{2,\mathrm{WL}}^{(2)}$ and
$C^{(2)}_\mathrm{WL}$ is that  no additional constant term is allowed on the right hand side of
\eqref{dualitywa}.  For the same reason,  the Wilson loop remainder function must then have
the same collinear limits as its amplitude counterpart,  i.e.
\beq
\label{abb}
  \cR_n^\mathrm{WL} \rightarrow \cR_{n-1}^\mathrm{WL} \ ,
\eeq
with no extra constant  on the right hand side of \eqref{abb}.

As is well-known \cite{dhks5}, the remainder function depends on the kinematics only through conformal cross-ratios.
In \cite{Anastasiou:2009kn}, the following definitions for the cross-ratios were adopted,
\beq
\label{ourcrossratios}
u_{i j} := {x_{i j+1}^2 x_{i+1 j}^2 \over x_{ij}^2 x_{i+1 j+1}^2}
\ ,
\eeq
where the $x_i$ are 't Hooft's region momenta, in terms of which momenta of on-shell scattered particles are defined as $p_i = x_i-x_{i+1}$.
The number of these cross-ratios is the same as that of two-mass easy box functions, i.e.~$n(n-5)/2$, if one ignores Gram determinant constraints \cite{Anastasiou:2009kn}. The inclusion of these constraints reduces this number to $3n-15$ \cite{dhks5}.  For $n=6$ one obtains three cross-ratios in both cases, whereas for octagon Wilson loops -- the case of principal interest for this paper -- one obtains 9 (12) cross ratios with(out) Gram determinant constraints.
In \cite{Anastasiou:2009kn} it was also shown that a generic cross-ratio, i.e.~one of the form $x_{ij}^2 x_{lm}^2 / ( x_{il}^2 x_{jm}^2)$ can be re-written as an appropriate product of the cross-ratios defined in \eqref{ourcrossratios}.

Finally, we briefly summarise the results of \cite{seven,Anastasiou:2009kn} on the
expected behaviour of the amplitude and Wilson loop remainder functions under triple-collinear limits.
We consider the case where the three momenta $p_1$, $p_2$ and $p_3$ become collinear,
\beq
\label{tripcol123sixpoints}
p_1 := x_1-x_2 = z_1P \  , \quad p_2 = x_2-x_3 = z_2 P \  , \quad p_3 = x_3-x_4 = z_3 P \ , \quad
z_1+z_2+z_3 =1 \ ,
\eeq
with $P^2=(p_1+p_2+p_3)^2=(x_1-x_4)^2 \to 0$.
In this limit, one finds \cite{seven,Anastasiou:2009kn} that
 \beq
 \label{ccc2}
 \cR_ n  \ \to \ \cR_{n-2} \, + \,  \cR _6(\bar{u}_1, \bar{u}_2, \bar{u}_3)
 \ ,
 \eeq
where
\beq
\label{collubars}
\bar{u}_1 =  {1 \over 1-z_3}{s_{12}  \over s_{123} } \ , \quad
\bar{u}_2 =  {1 \over 1-z_1}{s_{23}  \over s_{123} } \ , \quad
\bar{u}_3 = {z_1z_3 \over (1-z_1)(1-z_3)} \ .
\eeq

\section{Octagon Wilson loops}

\subsection{Definition of the kinematics}\label{sec:altkin}

We begin by reviewing briefly the special eight-point kinematics introduced in \cite{amoctopus}.
The starting point of \cite{amoctopus} is a ``zig-zag" configuration of momenta
embedded in $\mathbb{R}^{1,1} \subset \mathbb{R}^{1,3}$ where three of the
eight region momenta $x_i$ which determine the positions of the cusps are at infinity,
\begin{align}
\label{AMkinematics}
x_1 &= \left(0, 0\right)\, ,  \qquad
x_2 = \left({1\over 1 + \chi^+}, 0 \right)\, , \qquad
x_3 = \left({1\over 1 + \chi^+}, {1\over 1
    + \chi^-}\right)\, ,
 \\
 x_4 &= \left(1, {1\over 1 + \chi^-} \right)\, ,
 \qquad
 x_5 = \left(1, 1\right)\, ,
\qquad
x_6 = \left(\infty,     1\right)\, , \qquad
x_7 = \left(\infty, -\infty\right)\, , \\
x_8 &=
  \left(0, -\infty\right)\,  .
 \nonumber
\end{align}
This  configuration depends on
two  parameters, or more precisely lightcone cross-ratios $\chi^\pm$,
introduced in \cite{amoctopus}.
Alternatively one can combine them into a single complex parameter $m$ as follows,
\be
\chi^+ \, := \, e^{2\pi {\rm Im}(m)} \ , \qquad \chi^- \, := \, e^{-2\pi {\rm Re}(m)}\, .
\label{chi-m}
\ee
We can straightforwardly compute the 12 cross-ratios \eqref{ourcrossratios} arising from the
kinematics \eqref{AMkinematics}, obtaining
\beqa
\label{eightcross}
u_{15} &=& {\chi^{+}\over 1 + \chi^{+}} \ , \qquad u_{26} \ = \ {\chi^{-}\over 1 + \chi^{-}}
\ ,
\\ \nonumber
u_{37} &=& {1\over 1 + \chi^{+}} \ , \qquad u_{48} \ = \ {1\over 1 + \chi^{-}}\ ,
\\ \nonumber
u_{i\, i+3} &=& 1\ ,  \ \  \qquad \qquad i=1, \ldots , 8
\ .
\eeqa
One of the goals of this paper will  be to compare the strong-coupling
expression for the Wilson loop for the particular kinematics of \cite{amoctopus} with perturbative computations.
Direct use of the (1+1)-dimensional kinematics of \eqref{AMkinematics} is awkward for numerical evaluations,
due to the presence of infinities.  In order to circumvent this problem,
we will employ instead  three different (2+1)-dimensional sets of kinematic invariants
in our perturbative calculations, each of which is conformally equivalent to the set above, i.e.~gives rise to the same cross-ratios as in \eqref{eightcross}.  Since the Wilson loop remainder function is dual conformal invariant \cite{dhks5}, in practice
one can equivalently determine it with any of these three sets.
This also provides additional tests for our numerical calculations.

\noindent
{\bf Kinematics A}
\nopagebreak

\noindent
Our first choice of  (2+1)-dimensional kinematic variables is given by
\begin{align}
x_1 &= (1/2, 1/2, -1)\, , \qquad
  x_2 = \left({\chi^+\over 2 + 2 \chi^+}, {\chi^+\over 2 + 2 \chi^+}, -1\right)\, ,
  \nonumber\\
x_3 &= \left({1 + (2 + \chi^-) \chi^+\over
   2 (1 + \chi^- + \chi^- \chi^+)}, {-1 + \chi^- \chi^+\over
   2 (1 + \chi^- + \chi^- \chi^+)}, {-(1 + \chi^-) (1 + \chi^+)\over
    1 + \chi^- + \chi^- \chi^+}\right)\, , \nonumber\\
x_4 &= \left({1\over 2 + 2 \chi^-}, {-1\over 2 (1 + \chi^-)}, -1\right)\, ,\qquad
x_5 = (1/2, -1/2, -1),\nonumber\\
x_6 &= (-1/2, -1/2, 0)\, ,\qquad
x_7 = (0, 0, 0)\, ,\qquad
x_8 = (-1/2, 1/2, 0)\, ,
\label{8ptkinematics}
\end{align}
It can easily be shown  that  the corresponding cross-ratios \eqref{ourcrossratios} are
exactly as those in \eqref{eightcross}.

\noindent
{\bf Kinematics B}

\noindent
An even simpler  choice of kinematics,  which is  conformally equivalent
to that in \eqref{AMkinematics},  is characterised by taking the cross-ratios
of \eqref{eightcross} and supplementing them by eight Lorentz
invariants, all taken to be equal to minus one.
More precisely,  following \cite{Anastasiou:2009kn},   we start from  the full set of twenty momentum invariants
\beq
\label{20invariants}
x^2_{i\; i+2},x^2_{i+4 \; i+6}, \; x^2_{i \;i+3}, \; x^2_{i+4\; i+7},\; x^2_{i\;i+4} \ ,
\quad{i=1,\ldots,4}
\ ,
\eeq
and fix the following three- and four-particle momentum invariants as follows:
\beq
\label{other8}
x^2_{i+5 \; i+8}\ =\ -1\ =\  x^2_{i\;i+4} \ , \quad i=1,\dots,4\ .
\eeq
The remaining invariants are then determined in such a way that
the twelve cross-ratios are the same as in \eqref{eightcross}.
The computation then proceeds in terms of these invariants without
reference to specific region momenta.

\noindent
{\bf Kinematics C}
\nopagebreak

\begin{figure}
  \centering
  \scalebox{0.75}{
  \includegraphics[width=8cm]{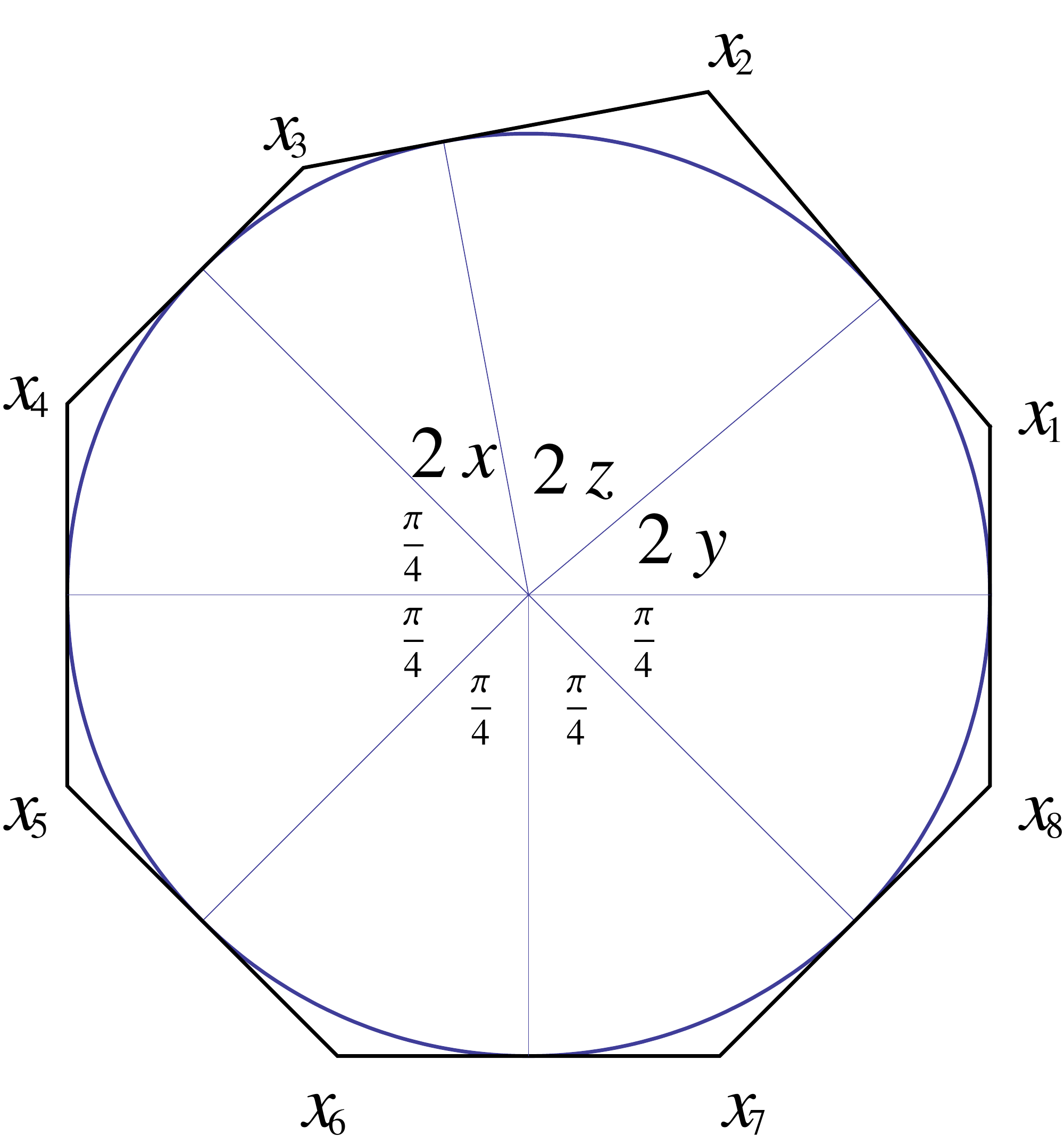}
  }
  \caption{\it Alternative kinematics, conformally equivalent to \eqref{8ptkinematics}, in which the spatial projection of
    the polygon circumscribes  the unit circle.   }
  \label{octagon}
\end{figure}

\noindent
The last (2+1)-dimensional kinematics we will use in the following  satisfies  the additional requirement \cite{amoctopus} that the
spatial projection of the polygonal contour circumscribes the unit
circle, see  Figure \ref{octagon}.%
\footnote{This kinematics is defined
  in terms of three angles $x,y$ and $z= 3 \pi/8 -x-y$. For this
  polygon to circumscribe a circle, the three angles must be
  positive. More generally, the kinematics in \eqref{altcoords} remains
valid for any values of $x$ and $y$ (even if the contour does not circumscribe
the unit circle).}
The coordinates of the region momenta are then given in terms of two parameters
$x,y$ as follows:
\begin{align}\nonumber
  x_1 &= (\tan y,1,\tan y) \ \  ,  \
x_2 = \left(-\tan z,\cos \left(z+2y\right) \sec z,\sin
    \left(z+2y\right) \sec z\right) \ , \\\nonumber
x_3 &= \left(\tan x,-\cos
    \left(x+\frac{\pi }{4}\right) \sec x,\sin \left(x+\frac{\pi
      }{4}\right) \sec x\right) \ , \\\nonumber
x_4 &= \left(-\tan \left(\frac{\pi
      }{8}\right),-1,\tan \left(\frac{\pi }{8}\right)\right) \ \ , \
x_5 =
  \left(\tan \left(\frac{\pi }{8}\right),-1,-\tan \left(\frac{\pi
      }{8}\right)\right) \ , \\\nonumber
x_6 &= \left(-\tan \left(\frac{\pi
      }{8}\right),-\tan \left(\frac{\pi }{8}\right),-1\right) \ \ , \
x_7 =
  \left(\tan \left(\frac{\pi }{8}\right),\tan \left(\frac{\pi
      }{8}\right),-1\right) \ , \\
x_8 &= \left(-\tan \left(\frac{\pi
      }{8}\right),1,-\tan \left(\frac{\pi }{8}\right)\right)\, ,
      \label{altcoords}
  \end{align}
where   $z=3\pi / 8 -x-y$.

The cross-ratios computed in this ``circular" kinematics coincide with
those of \eqref{eightcross},  if $x$ and $y$ are related to $\chi^\pm$ in the following way,
\begin{align}
  \chi^+&=\frac{-\tan x-\sqrt{2}+1}{\sqrt{2} \tan x-\tan x-1}\ , \\
  \chi^-&=\frac{-\tan y-\sqrt{2}+1}{\sqrt{2} \tan y -\tan y-1} \ .
\end{align}
The above equations can be easily inverted to give explicit analytic expressions for  $x$ (for $y$) as a function of
 $\chi^+$  ($\chi^-$).

Note that in the limit $m \to 0$ we have $\chi^\pm \to 1$,  and hence
\beqa
\label{cross-ratios-reg-oct}
u_{i\, i+4} ={1\over 2}, \quad i=1,\ldots,4\, ,
\\ \nonumber
u_{i\, i+3} =1, \quad i=1,\ldots,8 \, .
\eeqa
We will see in Section \ref{sec:5} that
these cross-ratios correspond to the case of a  regular octagon.
For the case of our  circular kinematics this is also obvious from Figure \ref{octagon},
since in this limit  $(x,y) \to (\pi/8, \pi/8)$.

\subsection{Results at strong coupling and   limits on the kinematics}
\label{scsection}

In the following we briefly review the strong-coupling result for the octagon remainder function in the special two-dimensional kinematics \eqref{AMkinematics}
introduced in \cite{amoctopus}.
This remainder function was found to be equal to  \cite{amoctopus}
\bea
\cR_8^{\rm AM}(m)&=&-\frac{1}{2} \log(1+1/\chi^+)\log(1+\chi^-) \nonumber\\
&&+\, \frac{7 \pi}{6} +
\int_{-\infty}^{\infty} dt \frac{|m| {\rm sinh} t}{{\rm tanh}(2t+2i \phi)}\log\left(1+e^{-2\pi|m|{\rm cosh}t}\right)
\, ,
\label{AMR8}
\eea
where the complex variable
\be
m := |m| e^{i \phi}
\ ,
\ee
is related to the lightcone cross-ratios
$\chi^\pm$ through \eqref{chi-m}.%
\footnote{Recall that these are related to standard
conformal cross-ratios via  \eqref{eightcross}.}
The expression for the octagon remainder function in  \eqref{AMR8}
is appropriate for the first quadrant of the complex $m$ plane ($\phi = 0, \ldots, \pi/2$), and is defined in the other quadrants by analytic continuation.   Furthermore, the remainder function is symmetric under
$\phi \to -\phi$ and $\phi \to \phi +\pi/2$, hence knowledge of this function in the first quadrant is sufficient.

For general $2n$-point  amplitudes (or $2n$-gon Wilson loops), the remainder function in the notation of  \cite{amoctopus}
can be represented as
\be
\cR_{2n}^{\rm AM} \,=\, \Delta A^{\rm BDS}_{2n} \,+\,
(A^{\rm sinh}_{2n} \,+\, A^{\rm extra}_{2n} )\, + \, A^{\rm periods}_{2n}
\, .
\label{2ngen}
\ee
The first term, $\Delta A^{\rm BDS}_{2n}:= A^{\rm BDS-like-even}_{2n}-A^{\rm BDS}_{2n}$, in the case of the
octagon gives rise to the log-log term in \eqref{AMR8};
the second term in \eqref{2ngen} gives $7 \pi/ 6$ plus the integral in \eqref{AMR8}, and finally
$A^{\rm periods}_{8}=0.$
We find it  convenient to separate the contribution arising from  regular $2n$-gons (i.e.~when $\chi^\pm \to 0$),
and define  $A^{\rm sinh}_{2n} := A^{\rm sinh}_{{\rm reg}\, 2n} + \Delta A^{\rm sinh}_{2n}$.
Summarising, for the octagon we have:
\bea
\Delta A^{\rm BDS}_{8} &\!=\!& -\frac{1}{2} \log(1+1/\chi^+)\log(1+\chi^-) \, ,
\label{delbds8}\\
A^{\rm sinh}_{{\rm reg-}8} &\!=\!&\frac{5\pi}{4} \, ,
\label{Asinhreg8}\\
\!\!\!\! \Delta A^{\rm sinh}_{8} +  A^{\rm extra}_8 &\!\!=\!\!& - \frac{\pi}{12} +
\int_{-\infty}^{\infty}\!\!\!dt \, \frac{|m| {\rm sinh} t}{{\rm tanh}(2t+2i{\rm Arg}(m))}\log\left(1+e^{-2\pi|m|{\rm cosh}t}\right)
\label{Asinh8}\\
A^{\rm periods}_{8} &\!=\!& 0\, .
\label{Aextra8}
\eea
We recall that  $A^{\rm periods}_{2n} = 0$ for regular polygons, while
$A^{\rm extra}_{2n}=0$ for odd $n$ \cite{amoctopus}.

Two limits of these expressions will be important for our discussion, and have an interesting geometrical interpretation:
$m \to 0$ and $|m| \to \infty$.
The first limit, $m \to 0$ (or $\chi^\pm \to 1$) corresponds to the case of the regular octagon.
Note that in this limit $\Delta A^{\rm sinh}_{8} + A_8^{\rm extra}\to 0$, as   shown in \cite{amoctopus}, so that
\be
m\to 0: \qquad \cR_{8}^{\rm AM}(m) \,\longrightarrow\, \cR_{{\rm reg-}8}^{\rm AM}\,=\,-\frac{1}{2} \log^2 2 \, +\, \frac{5\pi}{4}\, .
\label{AMreg8}
\ee
In the second limit, as $|m|$ goes to infinity, the cross-ratios $\chi^\pm$ either vanish or become
infinite. For example, in the limit where ${\rm Re}(m)\to +\infty$ and ${\rm Im}(m)\to -\infty$,
one has $\chi^+ \to 0$ and $\chi^- \to 0$.
In this case one can show that
\be
 \cR_{8}^{\rm AM}(m) \to \frac{5\pi}{4}-\frac{\pi}{12}=\frac{7\pi}{6}\, .
\ee
Intriguingly, this coincides with twice the value of the regular hexagon remainder function, $\cR_{6}^{\rm AM} = 7 \pi / 12$ \cite{amoctopus},
for which one has (cf.~\eqref{delbds8}--\eqref{Aextra8}):
\be
\cR_{6}^{\rm AM}\,=\,
\Delta A^{\rm BDS}_{6}\,+\,A^{\rm sinh}_{{\rm reg-}6}\,+\,\Delta A^{\rm sinh}_{8} \,+\,A^{\rm extra}_{6}\,=\,
0\,+\,\frac{7\pi}{12}\,+\,0\,+\,0 \, .
\label{AMhexreg}
\ee
The fact that $\cR_8 \to 2\cR_6$ in this limit has a general physical explanation, also valid at weak coupling.
In general, the  large-$|m|$ limits of $\cR_8$ correspond to various double-soft or soft-collinear limits
\cite{amoctopus}. In particular, the limit considered above, $\chi^\pm \to 0$, corresponds to a double-soft limit, i.e.~a situation where  two consecutive edges are becoming small.

This double-soft limit can be regarded as a special triple-collinear limit, where $(p_1, p_2, p_3) \to (z_1 P, z_2 P, z_3 P)$ and furthermore $z_{1,2}\to 0$ (as usual, $P := p_1 + p_2 + p_3$, $z_1 + z_2 + z_3 = 1$).
This implies that the eight-point Wilson loop  collapses to
a six-point Wilson loop plus the corresponding triple-collinear splitting amplitude.
As reviewed in Section \ref{review}, the latter is nothing but the six-point remainder function \cite{seven}, which for the special kinematics we are considering is just a constant.
Hence, we find that for $\chi^{\pm} \to 0$,  $\cR_8 \to \cR_6 +\cR_6$,
where the second $\cR_6$ is precisely the contribution of the triple-collinear splitting amplitude.
In summary we have,
\be
|m| \to \infty : \qquad \cR_{8}^{\rm AM}(m) \,\longrightarrow\, 2 \,\cR_{{\rm reg-}6}^{\rm AM}\,=\,
\frac{7\pi}{6}\, .
\label{2R6AM}
\ee
In order to visualise the strong-coupling remainder function \eqref{AMR8}, and to compare it with
our result for the two-loop remainder function,  we will plot both of them in the following section in
Figure \ref{ramagainst2loop} as functions of Re$(m)$ and Im$(m)$.

We shall now define a rescaled octagon remainder function at strong coupling,
which can be directly compared to our weak-coupling results in the next section.
To this end, we first shift $\cR_8^{\rm AM}(m)$  as defined in  \eqref{AMR8} by its value at $m=0$,
and then divide by the maximal variation
$\cR_8^{\rm AM}(0)-\cR_8^{\rm AM}(\infty)$.
Using the limits in  \eqref{AMreg8} and \eqref{2R6AM}, we get
\be
\overline{{\cR}}_8^{\,\rm AM}(m)\, := \, \frac{\cR_{8}^{\rm AM}(m) \,-\, \cR_{{\rm reg-}8}^{\rm AM}}
{\cR_{{\rm reg-}8}^{\rm AM}\,-\,2\cR_{{\rm reg-}6}^{\rm AM}} \,=\,
\frac{\cR_{8}^{\rm AM}(m) \,+\,\frac{1}{2} \log^2 2 \, -\, \frac{5\pi}{4}}{-\frac{1}{2} \log^2 2 \, +\, \frac{\pi}{12}}
\, .
\label{RAM8rescaled}
\ee

\subsection{The octagon remainder function at two loops}

\begin{figure}[h]
\begin{center}
\includegraphics[width=6cm]{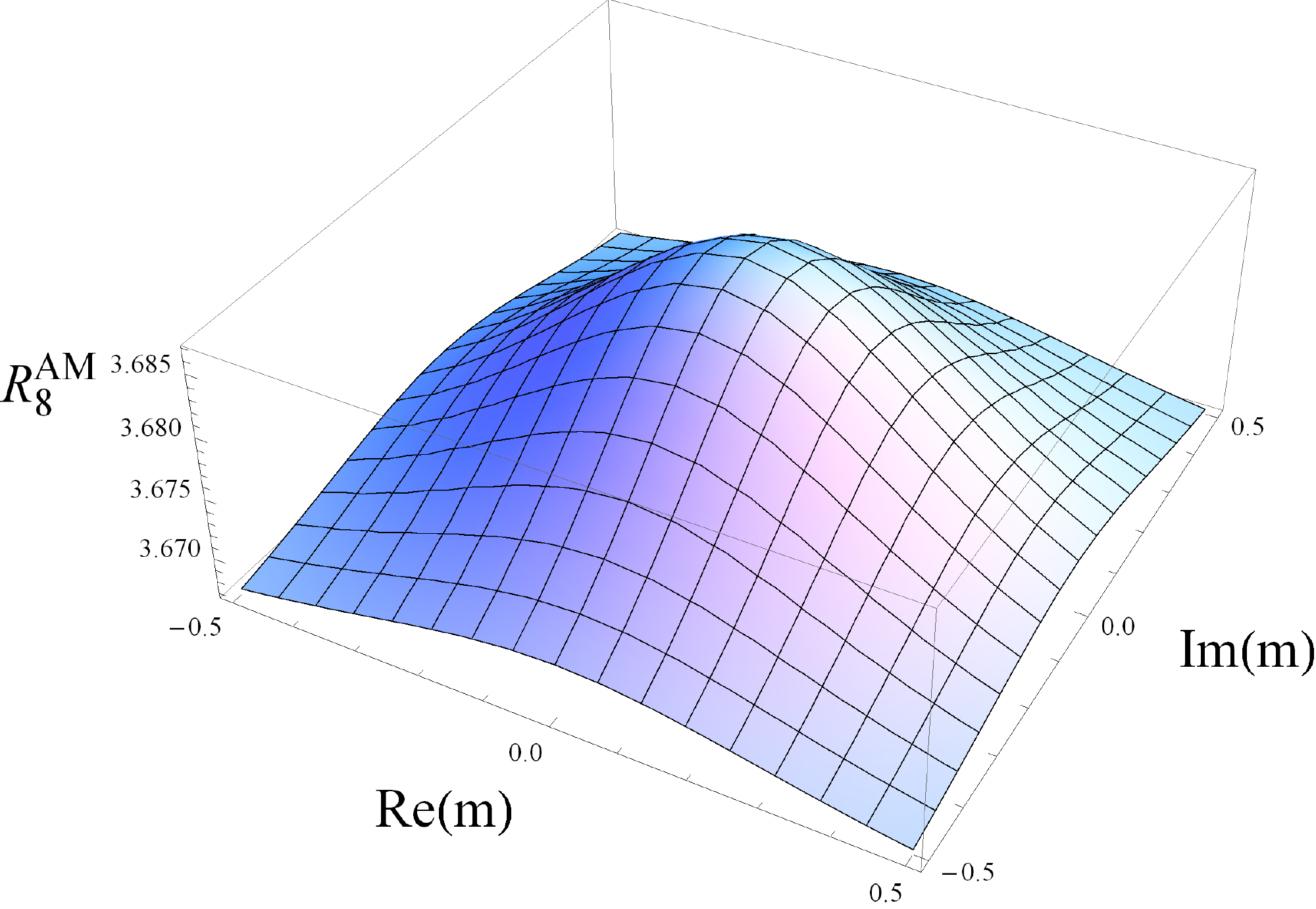}\qquad
\includegraphics[width=6cm]{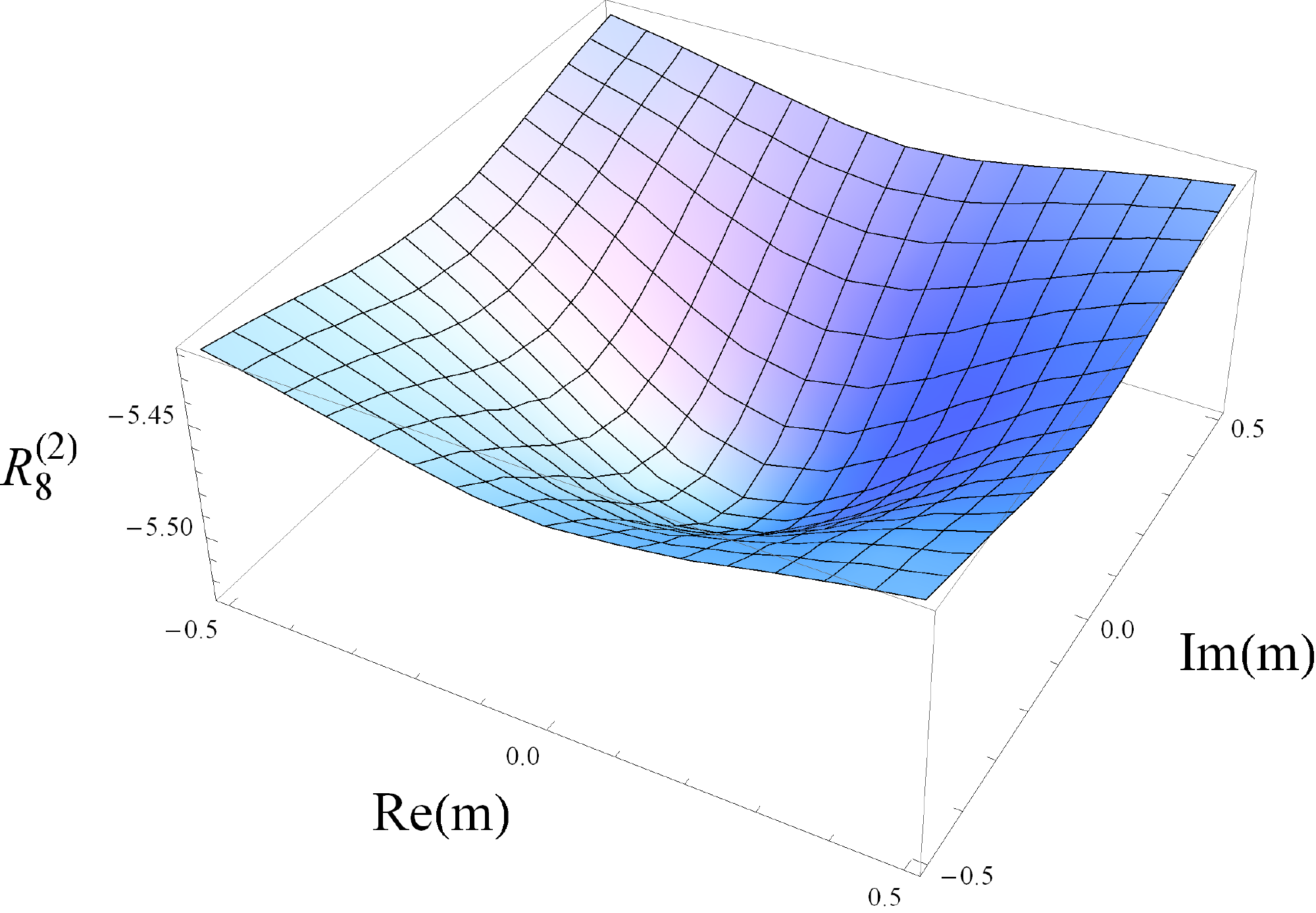}
\end{center}
\caption{\it The octagon remainder function in the special kinematics considered in  {\rm  \cite{amoctopus}}
as a function of ${\rm Re}(m)$ and ${\rm Im}(m)$.
The figure on the left represents the strong-coupling result $\cR_{8}^{\rm AM}$ derived in \eqref{AMR8},
while on the right we plot our two-loop result~$\cR_{8}^{\rm (2)}$.
}
\label{ramagainst2loop}
\end{figure}

We now present our results for the octagon remainder function  at weak coupling, and compare the perturbative remainder function to its strong-coupling counterpart.
We have computed the Wilson loop  at two loops using the general set-up developed in our earlier work \cite{Anastasiou:2009kn}, to which we refer the reader for a discussion of the numerical routines employed.
The remainder function is then extracted from the two-loop Wilson loop using  \eqref{wbds}.

 In order to compare our results to the remainder function evaluated at  strong coupling  we are free to use any conformally equivalent kinematic set, as the remainder function is dual conformal invariant \cite{dhks5}.
As mentioned earlier, for practical purposes we will use  three different choices of kinematic variables, A, B and C,
all of which are conformally equivalent to the kinematics of \cite{amoctopus}, as described in Section \ref{sec:altkin}.
As a preliminary question, one can check dual conformal invariance by comparing the value of the two-loop remainder function  $\cR_{8}^{\rm (2)}$ at a fixed set of cross-ratios, realised through  the three different  kinematics A, B and C.%
\footnote{This provides a useful check of the validity of our numerical routines.} We have confirmed in a large number of cases that this is indeed the case, within the numerical errors. In Table \ref{table1} below we quote one particularly important point, namely $m=0$, which corresponds to a regular octagon.%
\footnote{This data point will be used later in order to define the rescaled two-loop remainder function,  which can be directly compared to the corresponding strong-coupling result in \eqref{RAM8rescaled}.}

\begin{table}[ht]
\begin{center}
\scalebox{0.90}[0.90]
{
\begin{tabular}{|c||c|c|c|}
\hline
${\rm kinematics}$ &
A &
B &
C
\\
\hline \hline
$\mathcal{R}^\mathrm{(2)}_{\mathrm{reg}-8} $   &
$-5.5275 \pm 0.0008$      &
$ -5.52745\pm  0.0008$ &
$-5.5284 \pm 0.0007$
\\
\hline
\end{tabular}
}
\end{center}
\caption{\it In this table we list  our results for the Wilson loop remainder function corresponding to a regular octagon  contour. The cross-ratios corresponding to this case, quoted in \eqref{cross-ratios-reg-oct}, are realised using three different kinematics, namely the sets {\rm A, B, C}.}
\label{table1}
\end{table}

To begin with, we show on the right hand side of  Figure~\ref{ramagainst2loop} a two-dimensional plot of our
result for  $\cR_{8}^{\rm (2)}$ as a function of the two real variables ${\rm Re}(m)$ and ${\rm Im}(m)$.
For comparison  we show the strong-coupling remainder function on the left.
Already at this stage one can note an obvious similarity in the shape
of these two functions. In the following, we will quantify this similarity in a more precise way.

To address this point, we begin by making the preliminary observation that the geometrical interpretation of the
$m\to 0$ and $|m| \to \infty$ limits discussed  in Section \ref{scsection} is  obviously valid at strong as well as at weak coupling.
Thus, \eqref{AMreg8} and \eqref{2R6AM} hold also in perturbation theory (at any loop order):
\bea
|m| \to \infty  \!\!&:& \qquad \cR_{8}^{(2)}(m) \,\longrightarrow\, 2 \,\cR_{{\rm reg-}6}^{(2)}\, ,\\
m\to 0  \!\!&:& \qquad \cR_{8}^{(2)}(m) \,\longrightarrow\, \cR_{{\rm reg-}8}^{(2)} \, ,
\eea
and only the specific values for the regular octagon and the hexagon differ between the two regimes.
At two loops, we get for the regular hexagon%
\footnote{See Table 1 of \cite{Anastasiou:2009kn}, where the regular hexagon  case is obtained for all six-point cross ratios equal to~1.}
\be
\label{pi436}
\cR_{{\rm reg-}6}^{(2)} \, = \,   -2.707 \pm 0.002
\ .
\eeq
It was observed in \cite{Anastasiou:2009kn} that \eqref{pi436}  is compatible with
\beq
\cR_{{\rm reg-}6}^{(2)} \, \to
-\frac{\pi^4}{36} \, .
\label{2lhexr}
\eeq
For the regular octagon we will take as a reference value the following (see Table~\ref{table1}):
\be
\cR_{{\rm reg-}8}^{(2)} \, \simeq \, -5.5284\pm  0.0007 \, .
\label{2loctr}
\ee
In order to compare our weak-coupling result to  the strong-coupling calculation of \cite{amoctopus}, we introduce a shifted and rescaled remainder function, similarly to \eqref{RAM8rescaled} of the previous section,
\be
\overline{{\cR}}_8^{(2)}(m)\, := \, \frac{\cR_{8}^{(2)}(m) \,-\, \cR_{{\rm reg-}8}^{(2)}}
{\cR_{{\rm reg-}8}^{(2)}\,-\,2\cR_{{\rm reg-}6}^{(2)}}
\ .
\label{our8rescaled}
\ee
\begin{figure}[h]
\begin{center}
\scalebox{0.70}{
\includegraphics[width=12cm]{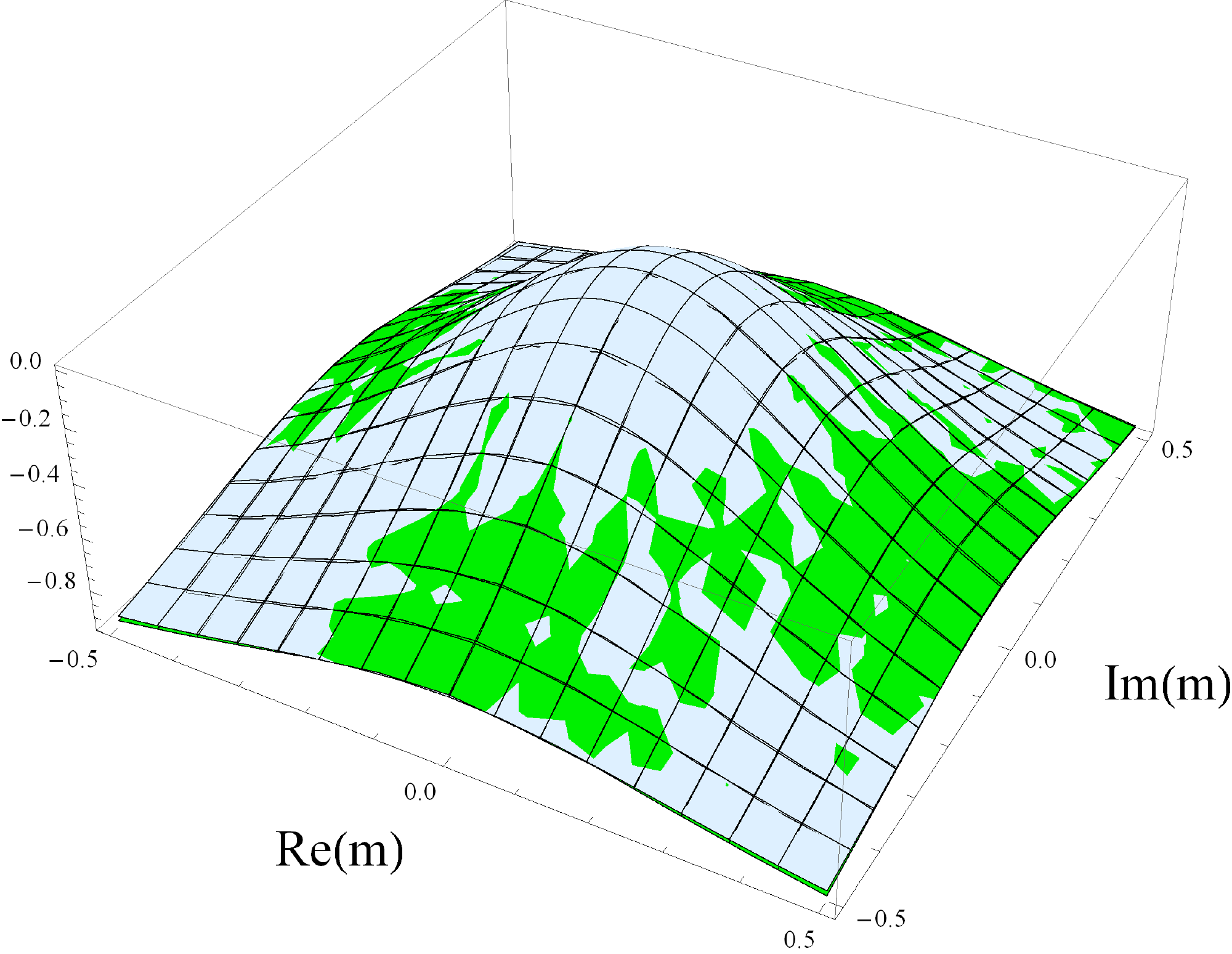}
}
\end{center}
\caption{\it The superposition of the 3D plots for
$\overline{{\cR}}_8^{\,\rm AM}(m)$ and  $\overline{{\cR}}_8^{(2)}(m)$.}
\label{ramagainst2looprescaled}
\end{figure}
In Figure~\ref{ramagainst2looprescaled}  we overlay the 3D plots of the rescaled
$\overline{{\cR}}_8^{\,\rm AM}(m)$ and  $\overline{{\cR}}_8^{(2)}(m)$
octagon remainder functions. Specifically, the latter was obtained through the evaluation of 121 data points using kinematics A, obtained by varying the cross-ratios $\chi^\pm$ as
\beqa
\chi^+ &=& e^{ \pi  (-1 + 2a / 10)}  \, , \quad a=0, \ldots , 10\, ,
\nonumber \\
\chi^- &=& e^{ \pi  (-1 + 2 b/10)} \, , \, \quad b=0, \ldots , 10\, .
\label{chivar}
\eeqa
A similar evaluation was also performed using kinematics C.
There appears to be an impressive numerical agreement between the rescaled remainder functions at strong and weak coupling.

\begin{figure}[h]
\begin{center}
{
\includegraphics[width=10cm]{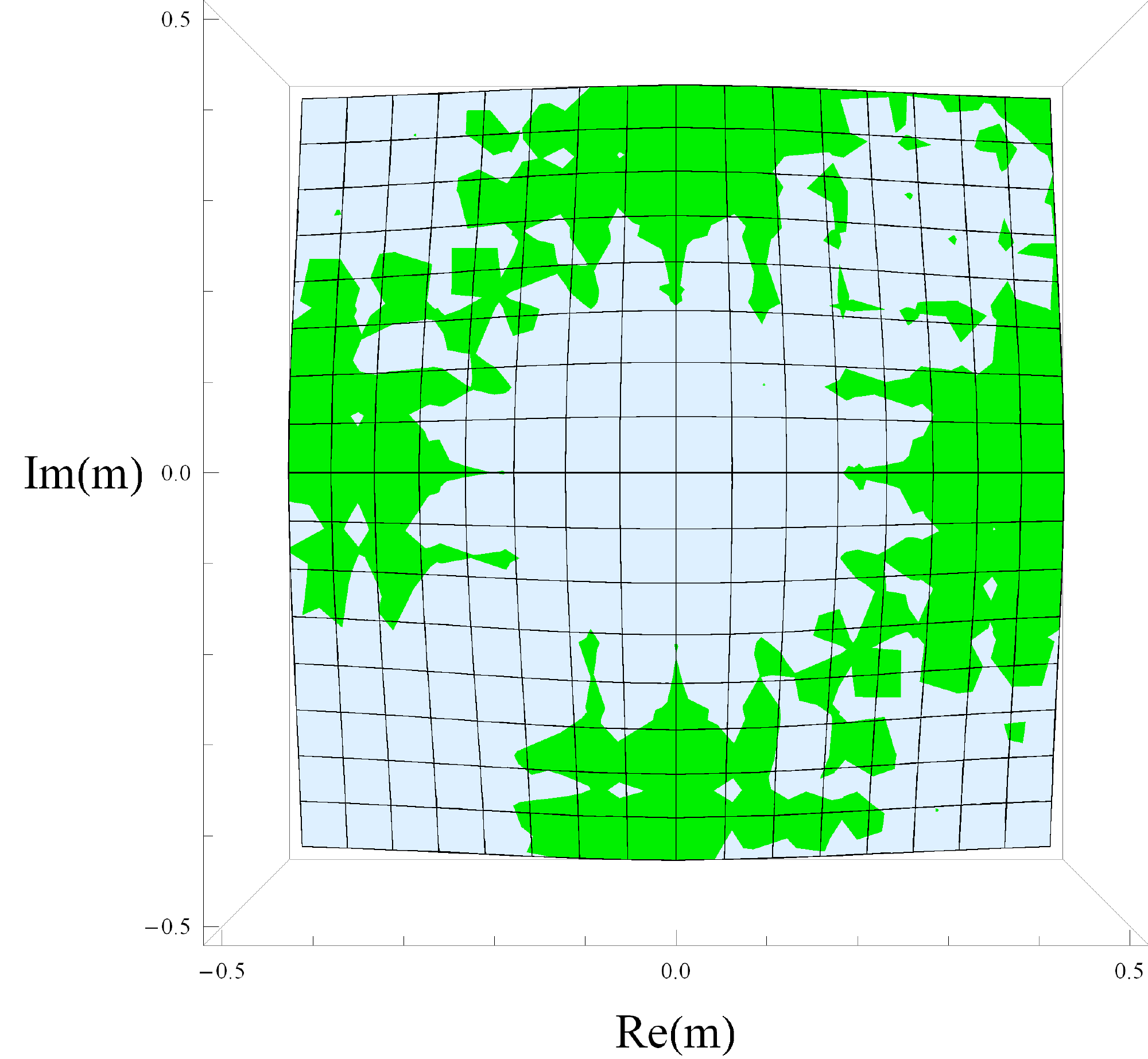}
}
\end{center}
\caption{\it A ``bird's eye view'' of Figure~\ref{ramagainst2looprescaled}
computed in kinematics {\rm A}.}
\label{3dplot2}
\end{figure}

Figure~\ref{3dplot2}  shows a ``bird's eye view'' of Figure~\ref{ramagainst2looprescaled}
computed using  kinematics A.
The green regions show where the two-loop rescaled remainder function dominates the corresponding strong-coupling result.
This could be viewed as an indication of a  numerical discrepancy between the two.
If this is a genuine (rather than purely numerical) discrepancy, it should be invariant under transformations generated
by the dihedral symmetry (cyclic permutations and parity) plus conformal transformations,%
\footnote{Conformal symmetry is necessary because the dihedral group alone transforms the twelve conformal cross-ratios
as well as the remaining eight kinematic Lorentz-invariants.
To undo the transformations on the latter while keeping the former
unchanged, one requires conformal transformations.}
 which manifest themselves \cite{amoctopus} as
$\phi \to -\phi$ and $\phi \to \phi+ \pi/2$.
The green region in Figure~\ref{3dplot2} is clearly invariant under one element of the group, namely
the reflection,  $\phi \to \pi/2-\phi$.
Actually, this symmetry is manifest for our numerical computation in kinematics A, one can easily check that it does not
involve conformal transformations and one is evaluating exactly the same integrals before and after the transformation.
However, the remaining symmetries do require conformal transformations of the kinematics and we can see that they
do not map the green region to itself exactly.
This indicates that the green region is not a genuine discrepancy.

\begin{figure}[h]
\begin{center}
\scalebox{0.70}{
\includegraphics[width=12cm]{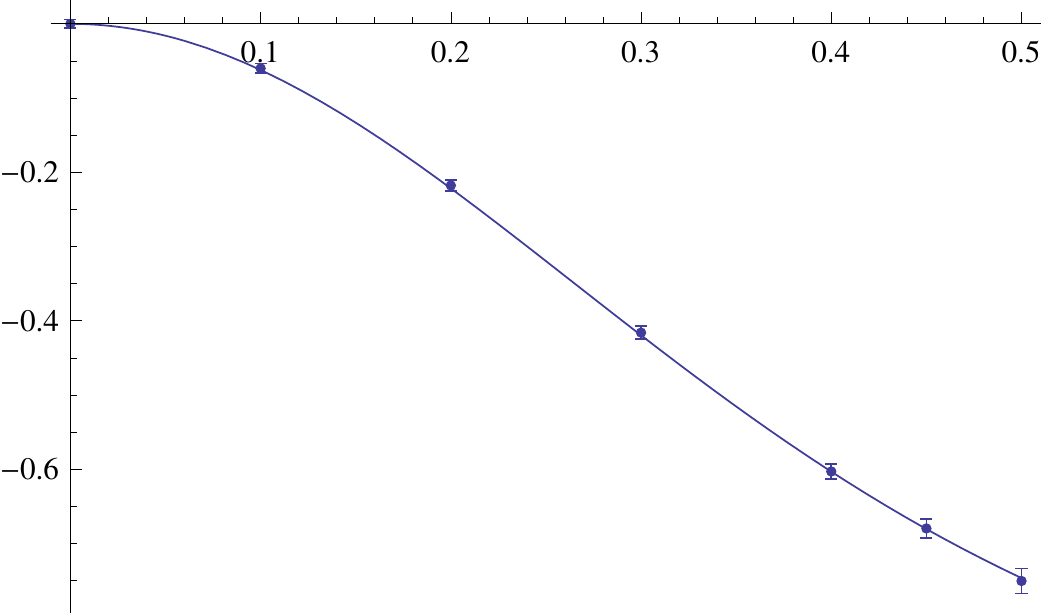}
}
\end{center}
\caption{\it This graph is a plot of $\overline{{\cR}}_8^{\,\rm AM}(m)$ and  $\overline{{\cR}}_8^{(2)}(m)$ as a
function of $|m|$ for $\phi=\pi/4$ computed in kinematics {\rm C}. }
\label{radialplotpi4}
\end{figure}
\begin{figure}[h]
\begin{center}
\scalebox{0.70}{
\includegraphics[width=12cm]{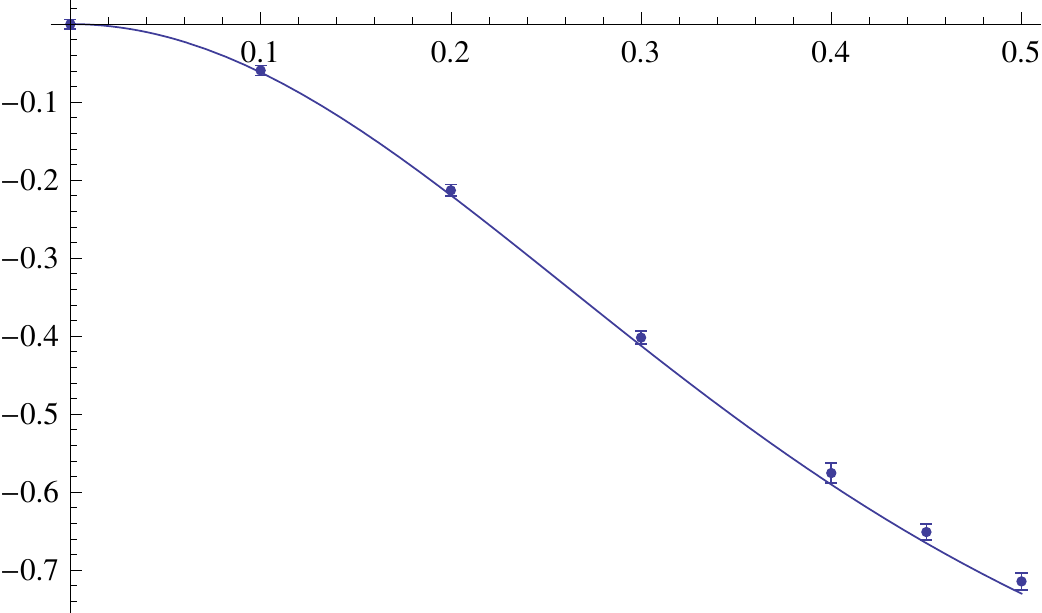}
}
\end{center}
\caption{\it Here we show  $\overline{{\cR}}_8^{\,\rm AM}(m)$ and  $\overline{{\cR}}_8^{(2)}(m)$ as a
function of $|m|$ for $\phi=0$ computed in kinematics C. We note that $\overline{{\cR}}_8^{\,\rm AM}(m)$ and  $\overline{{\cR}}_8^{(2)}(m)$ were defined to
coincide at $|m|=0$ and $|m|=\infty$.}
\label{radialplot0}
\end{figure}

We now proceed with a more detailed comparison of our weak-coupling results to the strong-coupling remainder function. Specifically, we find it useful to  consider two-dimensional  radial as well as polar plots.
In the radial plots,  we fix the argument of $m$ and vary $|m|$.  In Figure~\ref{radialplotpi4} we take  a diagonal slice,
$\phi=\pi/4$, whereas in Figure~\ref{radialplot0} we fix the argument of $m$ at $\phi=0$.
The weak- and strong-coupling results agree within the errors -- the mismatch between the two is barely visible
for both cases.
By comparing the two radial plots in Figures \ref{radialplotpi4} and \ref{radialplot0}, we note that the potential mismatch between the two depends on the value of $\phi$. To investigate this further we can zoom in on this mismatch at fixed values of $|m|$,  and plot the two remainder functions as functions of $\phi$.

\begin{figure}[h]
\begin{center}
\scalebox{0.70}{
\includegraphics[width=12cm]{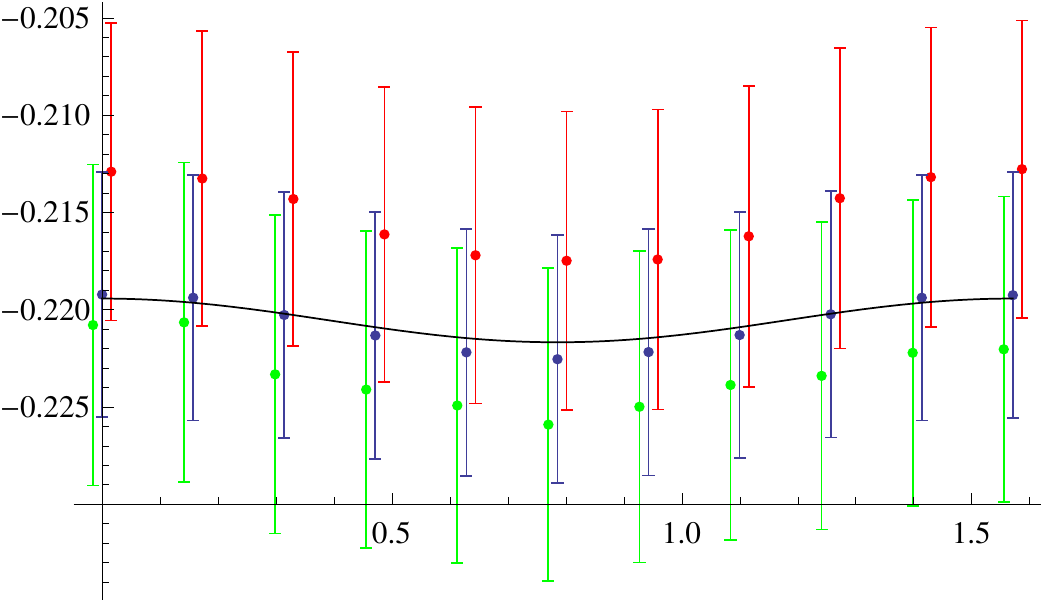}
}
\end{center}
\caption{\it Angular plot of $\overline{{\cR}}_8^{\,\rm AM}(m)$ (solid curve) and
$\overline{{\cR}}_8^{(2)}(m)$ for kinematic sets {\rm A} (blue), {\rm B} (green) and {\rm C} (red)
at $|m|=0.2$ as a function of $\phi$ in the first quadrant.
The two-loop results include the estimated error-bars. For clarity we have slightly shifted the green and red data points. }
\label{angular_0pt2_comparison}
\end{figure}

\begin{figure}[h]
\begin{center}
\scalebox{0.70}{
\includegraphics[width=12cm]{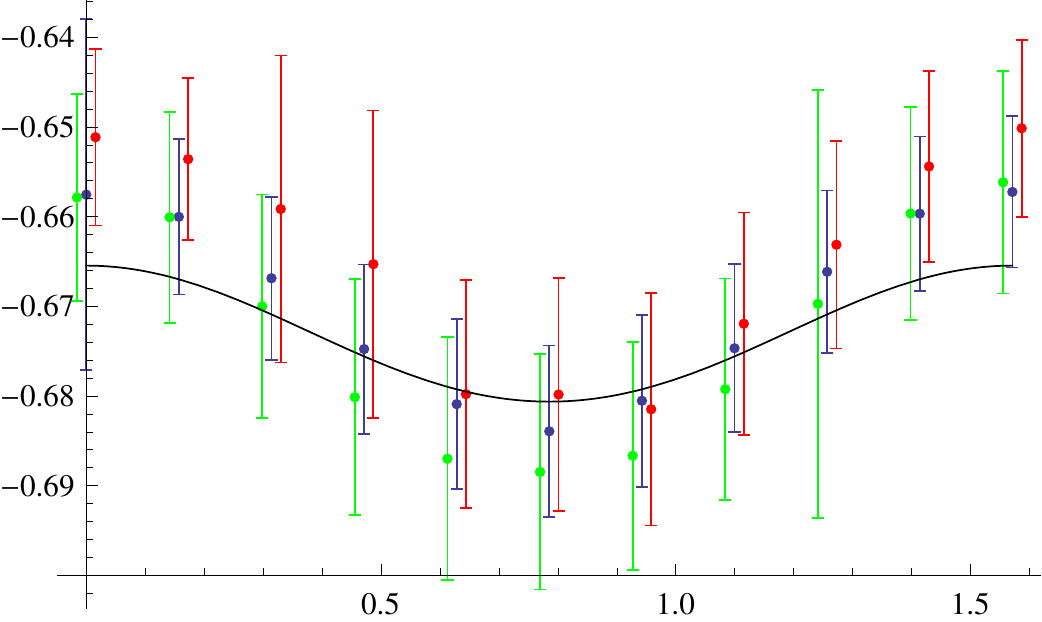}
}
\end{center}
\caption{\it Angular plot of $\overline{{\cR}}_8^{\,\rm AM}(m)$ (solid curve) and
$\overline{{\cR}}_8^{(2)}(m)$ for kinematic sets {\rm A} (blue), {\rm B} (green) and {\rm C} (red)
at $|m|=0.45$ as a function of $\phi$.}
\label{angular_0pt45_comparison}
\end{figure}

Figure~\ref{angular_0pt2_comparison} shows a polar plot at $|m|=0.2$ where we vary $\phi$ between $0$ and $\pi/2$.
The strong-coupling prediction $\overline{{\cR}}_8^{\,\rm AM}(m)$ is depicted alongside our two-loop data
$\overline{{\cR}}_8^{(2)}(m)$, obtained for the three kinematic sets A, B and C.
We notice that the mismatch between the three conformally equivalent sets for $\overline{{\cR}}_8^{(2)}(m)$
(which is a good indication of the limits of our numerical precision) is comparable with the mismatch between the
strong- and weak-coupling data, and the remainder functions at strong and weak coupling are in agreement
within the errors.
Figure~\ref{angular_0pt45_comparison} shows a different polar plot, obtained by fixing $|m|=0.4$, with similar features compared to
Figure~\ref{angular_0pt2_comparison}.

It is important to note that the overall shift between the curve at strong coupling and our perturbative data
is extremely  sensitive to the precise numerical value  for the regular octagon $\cR_{{\rm reg-}8}^{(2)}$ remainder function.
In particular, even the digit in the fourth decimal place affects the results in the two polar plots in  Figures \ref{radialplotpi4} and \ref{radialplot0}. For these figures, we have used the result quoted in \eqref{2loctr}.

To conclude this section, we compare the strong- and weak-coupling remainder functions for large  $|m|$.%
\footnote{We would like to thank Fernando Alday and Juan Maldacena for suggesting this additional test.}
In this region,  the integral appearing on the right hand side of  \eqref{AMR8} becomes exponentially small, and the strong-coupling remainder function approaches the constant value $7 \pi/6$, i.e.~twice the result for a hexagon, as discussed
in Section \ref{scsection}.
Specifically, we have   evaluated the two-loop remainder function at  $|m|=1$ and $|m|=1.5$, varying the  phase  $\phi$ from $0$ to $\pi/2$ using  kinematics A and C.
The numerical errors in this asymptotic region are larger than the maximal variation expected from the strong-coupling side, hence our tests are less stringent here. In particular, it becomes difficult to detect the variation in $\phi$ of the two-loop remainder function.   The average value  we found for $\mathcal{R}^{(2)}$ in the region  is fully consistent
with twice the result for the two-loop hexagon remainder.

\subsection{Universality of the octagon remainder function}

The agreement between the two shifted and rescaled remainder functions points at a linear relation between
\eqref{AMR8} and the weak-coupling remainder function, namely
\beq
\label{rhleoght}
\cR_{8}^{ (2)} (u_{ij})\, =\, A^{(2)}  \, \cR_{8}^{\, \rm AM}
(u_{ij})\, +\, {B}^{(2)} \ ,
\eeq
where $u_{ij}$ denotes the various cross-ratio, and  $A$ and $B$ are two constants that we now determine.
We can evaluate \eqref{rhleoght} at special values of $m$, in particular for $m=0$, corresponding to a regular octagon, and for $|m| \to \infty$, where $\cR_8 \to \cR_6$ at strong and weak coupling. Doing so, we obtain
\bea
A^{(2)} &=&  \frac{\cR_\mathrm{8-reg}^{(2)}  -2\, \cR_6^{(2)} }{  \cR_\mathrm{8-reg}^{\rm AM }  -2 \,\cR_6^{\rm AM} }\, , \\
{B}^{(2)} &=& 2 \, \frac{ \cR_6^{\rm AM } \,\cR_\mathrm{8-reg}^{(2)}  - \cR_6^{(2)} \, \cR_\mathrm{8-reg}^{\rm AM }  }{
2\, \cR_6^{\rm AM}  - \cR_\mathrm{8-reg}^{\rm AM }  }\, .
\eea
Using  \eqref{AMreg8} and \eqref{2R6AM}, as well as \eqref{2lhexr} and \eqref{2loctr},
one arrives at the following numerical estimates for $A^{(2)}$ and $B^{(2)}$,
\bea
\label{Aestnumer}
A^{(2)} \, = \,  -5.41\pm 0.03 \, , \qquad
{B}^{(2)} \, = \, 14.5 \pm 0.1
\, .
\eea
Note that the value for $A^{\rm (2)}$ quoted above is compatible with%
\footnote{
In fact, if the errors were actually smaller than what we quote in \eqref{Aestnumer},
the numerical estimate $A^{(2)} =  ( -5.5284 +{\pi^4\over 18 }) / ( -\frac{1}{2}\log^2 2 + \frac{\pi}{12})$
seems to be in pretty good agreement with $-\pi^4/18$. }
$-\pi^4/18  \equiv 2\,\cR_{{\rm reg-}6}^{\rm (2)}$.
It would clearly be of great interest if one could find a physical interpretation for the quantity $A$.%
\footnote{We will explain in Section \ref{secgenrem} that  the linear
  shift  ${B}$
should actually be absorbed into a
redefinition of the remainder function at strong coupling.}

\begin{figure}[h]
\begin{center}
\scalebox{1}{
\includegraphics[width=12cm]{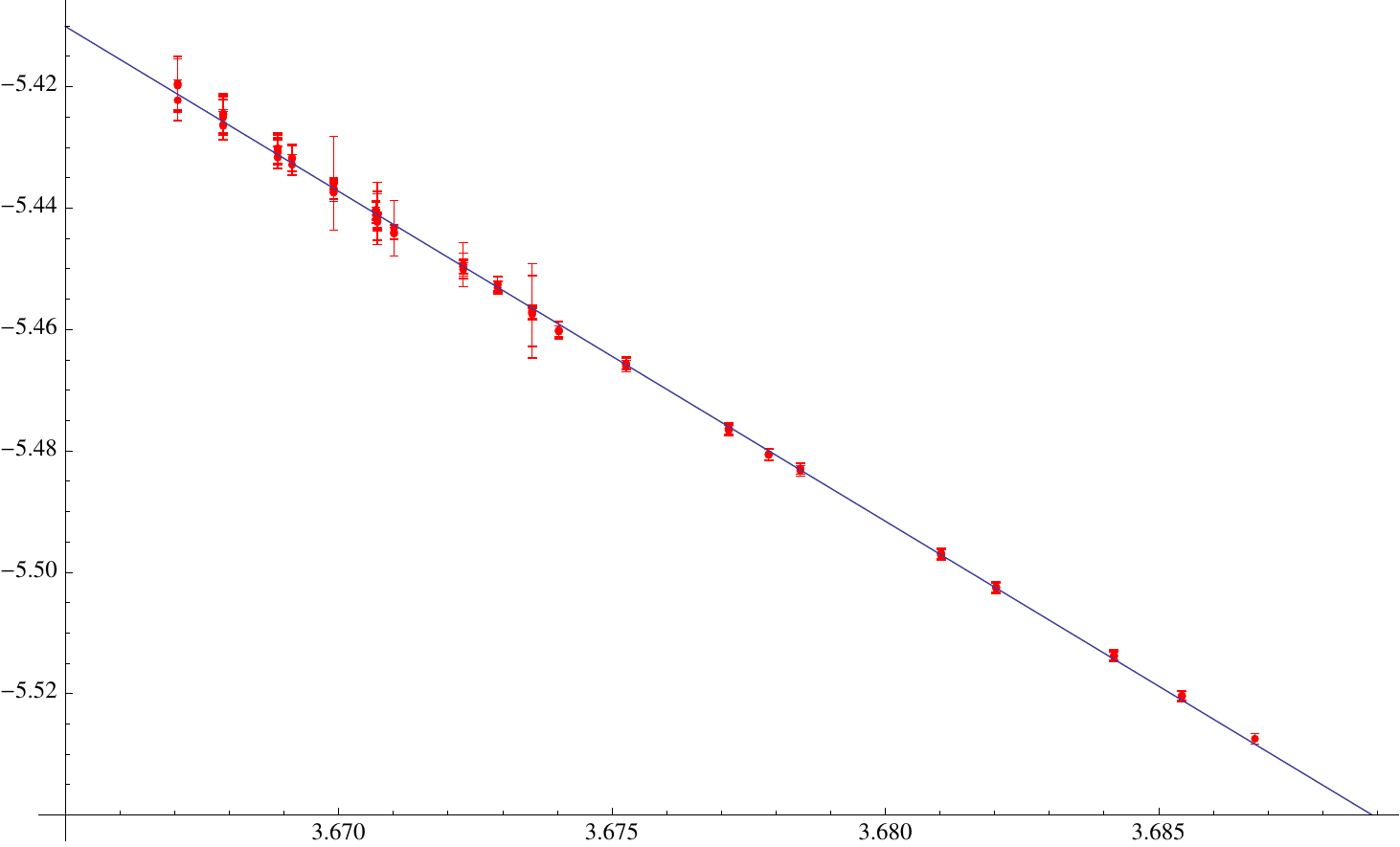}
}
\end{center}
\caption{\it The blue line represents  the least squares linear fit which maps the strong-coupling remainder function (on the horizontal axis) to our weak-coupling results (on the vertical axis). The numerical  data are also shown in red for comparison.}
\label{octagon_data_analysis}
\end{figure}

An alternative, statistically meaningful way to estimate the values
for the parameters $ A^{(2)}$ and $B^{(2)}$ consists in performing a
least squares fit of our numerical data to the corresponding
strong-coupling result calculated using \eqref{AMR8}.
We used a set of 121 data-points in kinematics A, obtained by varying the cross-ratios according to
\eqref{chivar}.
From this we obtained the following estimates:
\beq
\label{ABfit}
A^{(2)} \ = \ -5.434 \pm 0.012\, , \qquad
B^{(2)}\ = \  14.505 \pm 0.045 \ .
\eeq
This can be compared to the numerical values for these parameters in \eqref{Aestnumer}.
The values of $A^{(2)}$ and ${B}^{(2)}$ in \eqref{ABfit} and \eqref{Aestnumer} are in agreement within
numerical errors.
The non-reduced $\chi^2$-square of our fit is $\sim 31$, to be compared to the number of degrees of freedom, which in our case is%
\footnote{We recall that the number of  degrees of freedom is  the number of data points minus number of estimated parameters.} 119.
This is a rather small result, pointing at a possible overestimate of our error,
still confirming that our data are compatible with the strong-coupling result of \cite{amoctopus}.
In Figure \ref{octagon_data_analysis} we present the least squares linear fit together with our numerical results.

The errors as quoted by our numerical evaluation routines are usually larger  than the discrepancy between the different values of the remainder function obtained for a fixed set of cross-ratios but with conformally equivalent kinematics.  We believe that this discrepancy across conformally equivalent sets, which we found to be roughly of the order of $10^{-3}$, is a slightly more accurate way to quantify (and in fact reduce) our errors, although perhaps too optimistic across the whole set of data points. Of course, a $\chi^2$-test performed with smaller errors is more stringent.
For example, by assuming that the numerical error is roughly $1/2$ the error as quantified by our routines, the new
$\chi^2$-square becomes roughly equal to the number of degrees of freedom.

To summarise, the numerical data we have obtained  lead us to conjecture that the shape of the remainder function
for the octagon Wilson loop is universal. In the following section we will further sharpen the
statement of shape-invariance of the remainder function,  and apply it to generic $2n$-gon Wilson loops.

\section{Remainder functions for general $2n$-gons}
\label{secgenrem}

Here we will discuss the shape-invariance of the remainder functions in terms of the original
unrescaled and unshifted remainders for general polygonal Wilson loops (with an even number of edges $2n$).

First of all, what do we mean by  shape-invariance, or  invariance of the functional form  up to shifts and rescalings?
Clearly this invariance (if it persists for general $n$ and for general kinematics) implies a
linear relation between the remainder function at strong coupling and its perturbative expressions at each loop order $L$:
\beq
\label{rhl}
\cR_{2n}^{ (L)} (\{u_{ij}\})\, =\, A_{2n}^{ (L)} \, \cR_{2n}^{\, \rm strong} (\{u_{ij}\})\, +\, B_{2n}^{ (L)}\ ,
\eeq
where $A_{2n}^{ (L)}$ and $B_{2n}^{ (L)}$ are constants.
We  assume  here that the remainder functions at strong and weak coupling have the correct normalisation under simple
collinear limits,
i.e.~\cite{seven,Anastasiou:2009kn}
\beq
\label{cl}
\cR_{2n} \to \cR_{2n-1} \, ,
\eeq
where no constant term can appear on the right hand side of \eqref{cl}.
By taking simple collinear limits of \eqref{rhl}, one obtains the consistency condition
\beq
A_{2n}^{ (L)} \, \cR_{2n-1}^{\, \rm strong} \, +\, B_{2n}^{ (L)} \ = \
A_{2n-1}^{ (L)} \, \cR_{2n-1}^{\, \rm strong} \, +\, B_{2n-1}^{ (L)} \ .
\eeq
From this relation, one concludes that $A$ and $B$ must be $n$-independent.
Next, we consider triple collinear limits of \eqref{rhl}, and using \eqref{ccc2}
we arrive at the consistency condition
\beq
\label{ab}
A_{2n}^{ (L)} \, \cR_{2n-2}^{\, \rm strong} \, +\,  A_{2n}^{ (L)} \, \cR_{6}^{\, \rm strong} \, + \, B_{2n}^{ (L)}
 \ = \
A_{2n-2}^{ (L)} \, \cR_{2n-2}^{\, \rm strong} \, +\, B_{2n-2}^{ (L)} \, + \,   \cR_{6}^{{(L)} } \ .
\eeq
Importantly, in the particular two-dimensional kinematics considered in \cite{amoctopus}, the final six-point configuration has
no free cross-ratios left, i.e.~the six-point remainder function is just a constant.
Equation \eqref{ab}  implies immediately that $A$ is $n$-independent, in agreement with simple collinear limits.
Furthermore,  re-expressing  $\cR_{6}^{{(L)} }$ in terms of the corresponding strong-coupling result using \eqref{rhl},
we arrive at
\beq
B_{2n}^{ (L)} \, = \, B_{2n-2}^{(L)} \, + \, B_6^{(L)}
\ ,
\eeq
and hence $B_{2n}^{ (L)}  = (n-2) B_{6}^{ (L)}$. This result is clearly inconsistent with the result derived earlier that $B_{2n}^{(L)}$ is also $n$-independent unless $B_6^{(L)}=0$. In turn, this implies that $B_{2n}=0$ for all $n$.

In conclusion, simple and multi-collinear limits show that the weak-coupling result
for the remainder function and its strong coupling counterpart must be proportional,
\beq
\label{rhl2}
\cR_{2n}^{ (L)} (\{u_{ij}\})\, =\, A^{ (L)} \, \cR_{2n}^{\, \rm strong} (\{u_{ij}\})\ ,
\eeq
where $A$ depends on the coupling constant but not on the number of particles. No constant shift is allowed on the right hand side of
\eqref{rhl2}.

We found earlier in \eqref{rhleoght} that a nonzero value for ${B}$ was required in order to match the weak-coupling result
for the eight-point kinematics and the strong-coupling result \eqref{AMR8} of \cite{amoctopus}.
There is no contradiction between this and \eqref{rhl2}; indeed, as already pointed out in \cite{amoctopus},
on the right hand side of  \eqref{AMR8} there is room for an additive constant, which was not determined in that paper.
Setting
\beq
\label{badc}
\cR_{8}^{\, \rm strong}\, = \, \cR_{8}^{\, \rm AM} + C \ ,
\eeq
and using our earlier result $\cR_8^{(2)} = A^{(2)} \cR^{\, \rm strong}_8 + B^{(2)}$,
together with \eqref{rhl2} and \eqref{badc},  we immediately find $C = B^{(2)}/A^{(2)}$.

Notice also that consistency with multi-collinear limits requires that
\beq
\cR_{6}^{\, \rm strong}\, = \, \cR_{6}^{\, \rm AM} + {C\over 2} \ ,
\eeq
where we have also used that  $\cR_{8}^{\, \rm AM} \to 2  \cR_{6}^{\, \rm AM}$ under  a triple-collinear limits.

Finally, we note that, at first sight, it appears that the universality of the remainder function implies the loss of the
maximal  transcendentality principle for the remainder function
$\cR_{2n}^{(2)}$.
This is because  the strong-coupling octagon remainder function  \eqref{AMR8} does not have a uniform degree of transcendentality in the first place,
and rescalings are not sufficient to  restore it.
However,  one should bear in mind that the kinematics we have been investigating is rather special (the remainder function in this
kinematics lives on a two-dimensional subspace of the full twelve-dimensional space of cross-ratios)
and this might obscure the transcendentality properties of the remainder function.
As a concrete example,%
\footnote{Many thanks to Zvi Bern for pointing out this example and a very useful discussion of how such loss of transcendentality
for special kinematics occurs in certain two loop integrals.}
one may consider the function $\log (1+x) / x$, which  appears to have transcendentality equal to zero as $x\to 0$.
This interesting issue clearly deserves further study.
It would also be very important to investigate the validity of the universality conjecture for different number of particles
and arbitrary kinematics,  and confront it with analytic computations at two loops.

\section{Regular $2n$-gons}\label{sec:5}

In this section we will discuss first the particular kinematics associated to regular polygons with $2n$ edges, and we will then move on to summarise the results of our numerical analysis we have performed on  Wilson loops with up to $30$ edges.
\subsection{The regular polygon kinematics}

Here we define the kinematic variables we use for the regular $2n$-gon Wilson loop.
Consider first the projection of the Wilson loop on the $(X,Y)$ plane. This projection defines a regular polygon and the positions of the vertices are given by $Z_{k} = \exp ( i 2 \pi k  / 2n)$, where $Z:= X+ iY$.
In the time direction, the momenta  proceed following a zig-zag pattern, which leads to the following positions for the vertices
(in the notation $(t, Z)$):
\beq
x_{2k} \ = \ \big( 2 \sin {\pi \over 2n} , \, e^{ i \pi {2k+1\over n}}\big)\ , \qquad
x_{2k+1} \ = \ \big( 0 , \, e^{ i \pi {2k\over n}}\big)\ .
\eeq
It is immediate to check that $ (x_{2k} - x_{2k+1})^2 = 0$.
The expression of the  various Mandelstam variables needed in the calculation are given by
\beqa
\nonumber
x_{2k \, 2j}^2 &=& -4 \sin^2 {k-j\over n} \pi
\ ,
\\ \nonumber
x_{2k +1\, 2j+1}^2 &=& x_{2k \, 2j}^2 \ ,
\\
x_{2k \, 2j+1}^2 &=& 4 \Big( \sin^2 {\pi \over 2 n}  - \sin^2 {k- j -1/2 \over n } \pi \Big)
\label{regpolkin}
\eeqa
At six points,  there are only three independent cross-ratios in \eqref{ourcrossratios} given by \cite{dhksbum,dhks6}
\beq
\label{sixptcr}
u_{36}  \, = \,  {x_{31}^2 x_{46}^2 \over x_{36}^2 x_{4 1}^2} \, := \,  u_1 \ , \qquad
u_{1 4} \, = \,   {x_{15}^2 x_{24}^2 \over x_{14}^2 x_{2 5}^2} \, := \,  u_2
\ , \qquad
u_{25} \, = \,  {x_{26}^2 x_{35}^2 \over x_{25}^2 x_{3 6}^2} \, := \,  u_3
\ .
\eeq
It is immediate to check that for a regular hexagon $u_1 = u_2 = u_3 = 1$.
For a regular octagon, there are  four cross-ratios of the type $u_{i i+4}$, and eight of the type $u_{i i+3}$.  A short calculation shows that, in that case,
\beq
u_{i i+4} \ = \ {1\over 2} \ , \qquad
u_{i i+3} \ = \ 1  \ .
\eeq
Finally, we consider a generic $2n$-gon configuration. We find that there are only two possible results for the cross-ratios $u_{ij}$  depending on whether  $i-j$ is even or odd. Our result is
\beqa
\label{regpolcrra}
u_{ij} & = & 1 \ ,   \qquad \qquad \qquad \qquad \ i\, - \,j  \, = \, {\rm odd} \ ,
\nonumber \\
u_{ij} & = & 1\, - \, \left(
{
\sin {\pi\over   n}
\over
\sin {\pi a \over   n}
} \right)^2
\ , \qquad
i\, - \,j  \, = \, 2a \ ,
\eeqa
where $a$ is an integer.
\subsection{Results at strong and weak coupling}
At strong coupling, Alday and Maldacena in~\cite{amoctopus} considered
the area of the minimal surface  associated with a regular
$2n$-gon. In particular they computed the $A^{\rm sinh}_{2n}$
contribution (in the notation of \eqref{2ngen}) to the remainder function:
\beq
\label{regstro}
A^{\rm sinh} _{2n}\ = \ \pi \, \left( {3\over 4} n \, - \, 2 \, + \, {1\over n }\right)
\ .
\eeq
They also considered the $n\to \infty$ limit, where the polygon tends
to a circle. In this case, the area develops a divergence proportional
to the perimeter of the loop, and the renormalised area is
\beq
A^{\rm ren} \ = \ - 2 \pi\ .\label{Acirc}
\eeq
One may wonder if the corresponding result for the remainder function at weak coupling shares any of the features of
\eqref{regstro}.

With the numerical approach outlined in the previous sections, we have
computed the (complete) remainder function for regular
polygons with up to $30$ edges at two loops in perturbation theory.
The regular Wilson loop was calculated in the kinematics
\eqref{regpolkin}, and the remainder function was obtained by
subtracting the BDS ansatz. The remainder function depends on $n$
through the regular polygon cross-ratios in \eqref{regpolcrra}.
In our perturbative computation we have made no attempt to isolate
the $A^{\rm sinh}$ contribution and our results below will always
correspond to the complete  remainder function.

Our data are recorded in the Table \ref{table2} below,
and show quite clearly that, at large $n$, $\cR^{(2)}_{2n} \sim n$
reproducing  the behaviour of the strong-coupling result \eqref{regstro}.

\begin{figure}[h]
\begin{center}
\includegraphics[width=10cm]{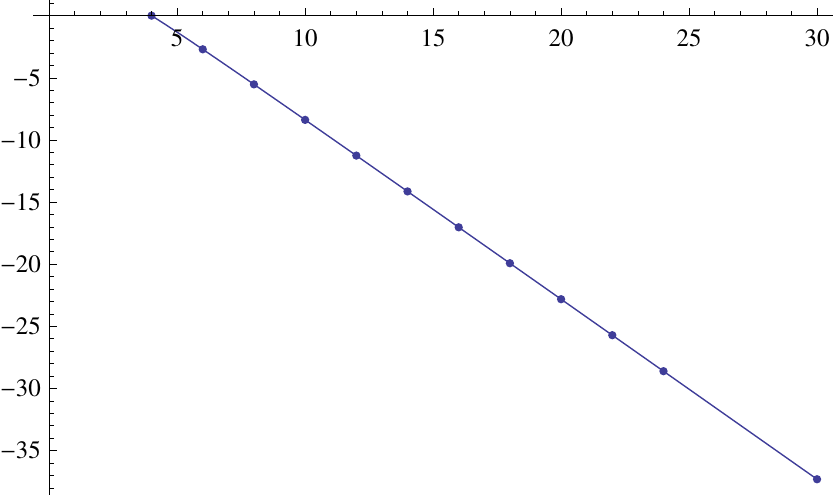}
\end{center}
\caption{\it The data points of the regular $2n$-gon remainder
are well approximated by the fit \eqref{fit1}. The horizontal axis
gives the number of edges
$2n$ from 4 to 30.}
\label{npoint_regplot}
\end{figure}

\begin{table}[ht]
\begin{center}
\scalebox{0.85}[0.85]{
\begin{tabular}{|c||c|c|c|c|c|c|c|c|c|c|}
\hline $2n$ & $8$ & $10$ & $12$ & $14$ & $16$ & $ 18 $ & $20$& $22$ & $24$ & $30 $
\\
\hline \hline $\mathcal{R}^\mathrm{(2)}_{2n} $   &  -5.528  &  -8.386 & -11.262 & -14.145 & -17.035 &-19.926 & -22.821& -25.717& -28.614& -37.311
\\
\hline
\end{tabular}
}
\end{center}
\caption{\it Numerical values for the Wilson loop remainder function corresponding to a $2n$-gon contour.
For a four-edged loop the remainder function vanishes, as  there are no non-trivial cross-ratios at four points.
For the six-point regular polygon, all the cross-ratios are equal to one, and the corresponding value of the remainder function is given in {\rm \eqref{pi436}}, see also
{\rm \eqref{2lhexr}}.
 }
\label{table2}
\end{table}
We have also performed a fit of our data to  a linear ansatz
$\mathcal{R}^{(2)}_{\rm fit,1}  (n) =r_1 n    + r_0 + r_{-1} /n $.
The result of such a fit,  using all the data points  in Table
\ref{table2},  gives
\beq
\mathcal{R}^{(2)}_{\rm fit,1}  \ = \ \pi \,
\Big (
- 0.9255 \, n \, + \, 2.026 \, -\,  { 0.346\over n}
\Big)
\ .\label{fit1}
\eeq
Note that $n$ is half the number of edges.%
\footnote{In \eqref{fit1} and \eqref{fit2} below we quote the result up to the first digit affected by the error.}
In Figure~\ref{npoint_regplot}
we compare the data points with the fit \eqref{fit1}.
The difference between the fit and the data is $\lesssim 0.01$.

To see how robust the fit is, we have also considered a slightly more
general fit with an additional $1/n^2$ term, and found
\beq
\mathcal{R}^{(2)}_{\rm fit,2}  \ = \ \pi \,
\Big (
-0.9239 \,n \, + \, 1.9955 \, -\,  \frac{0.181}{n} \, - \, \frac{0.228}{n^2}
\Big)
\ .\label{fit2}
\eeq
The difference between this fit and the data is $\lesssim 0.0004$. One could also perform fits including  higher inverse powers of $n$, however these do not change our conclusion about the linear behaviour of the data as a function of $n$ at large $n$ and the estimate for the slope of the line.

A few comments are in order.
Firstly, we note  that at large $n$ the $n$-dependence of $\mathcal{R}^{(2)}$ is the
same as that of $A^{\rm sinh}$ at strong coupling -- they both grow
linearly in $n$. Secondly, we observe  that the constant term is well
approximated by $2 \pi$ and up to a sign is the same as the strong
coupling result in \eqref{regstro}. We have already
noticed earlier a similar sign swap between strong and weak coupling results, see Figure
\ref{ramagainst2loop}.

Here we will not attempt to make a precise
connection between these results and the universality of the functional
form of the remainder function advocated in earlier sections.
To do so,  we would require the full expression for the remainder function at
strong coupling. In addition to $A^{\rm sinh}$, given in \eqref{regstro}, we would also need
expressions for  $\Delta A^{\rm BDS}_{2n}$ and $A^{\rm extra}_{2n}$  (which can be obtained in principle from \cite{amoctopus} but  are not known explicitly for
arbitrary $n$).
In addition to these, as already noted in \cite{amoctopus}, there is a
regularisation-prescription dependent shift. This shift is a kinematic-independent constant, which does depend on $n$.
Therefore, for regular polygons it is not immediately straightforward to give a direct relation between the complete
remainder functions at strong and weak coupling.

\section{Conclusions}

In this paper we have presented strong evidence in favour of a
remarkable  albeit unexpected agreement between the
remainder functions computed at strong and at weak coupling.
We would like to comment briefly on the conclusions  one can draw from this.

{\bf 1.}
Our evaluation of the two-loop remainder function is mostly numerical. Therefore there is always  the
possibility that there exists a genuine disagreement   between the
strong- and weak-coupling remainder functions.
If this is the case, it is still remarkable that it must be very small, as quantified in this paper.

{\bf 2.}
The second possibility is of course that there is a precise agreement
between the remainder functions at strong and weak coupling. Even if
this holds only for eight points, and for the special kinematics of
\cite{amoctopus}, this would be a highly non-trivial result, but it is
more natural to conjecture that it holds generally
for any number of points, and for arbitrary kinematics.
In this case, the  remainder function could be cast in the form
\beq
\label{boldconjecture}
\cR (a)  \, = \, A (a) \cR_{\rm strong}
\ ,
\eeq
where  the function $A (a) $ has a perturbative expansion starting  at two loops and is independent of $n$ and of the kinematic variables
(which are all contained in $\cR_{\rm strong}$).
If \eqref{boldconjecture} is correct, it would be of great importance to understand the physical meaning of the quantity
$A(a)$.

The conjecture of universality for the Wilson loop in
\eqref{boldconjecture}, would imply novel
non-renormalisation theorems or symmetry arguments.
An exciting way forward would be to develop analytic methods for
computing Wilson loops at strong coupling for other values of $n$
and/or in different kinematics, perhaps by extending the results of
\cite{amoctopus}.
Our conjecture of universality of the Wilson loop remainder function could then be further tested in these cases.

\vspace{1cm}
\section*{Acknowledgements}

It is a pleasure to thank  Fernando Alday,  Zvi Bern, Lance Dixon, Juan Maldacena and Bill Spence
for useful discussions and comments on a preliminary version of this paper.  We are especially grateful to Babis Anastasiou and Bill Spence for collaboration
on the precursor  to this work.
The numerical evaluations carried out in this paper would not have been possible without the codes
developed by Babis Anastasiou for \cite{Anastasiou:2009kn}.
This work was supported by the STFC under the Queen Mary Rolling Grant
ST/G000565/1 and the IPPP Grant ST/G000905/1.
VVK acknowledges a Leverhulme Research Fellowship and GT is supported
by an EPSRC Advanced Research Fellowship EP/C544242/1 and by an
EPSRC Standard Research Grant EP/C544250/1.


\vspace{1cm}

\begin{thebibliography}{99}




\bibitem{abdk}
  C.~Anastasiou, Z.~Bern, L.~J.~Dixon and D.~A.~Kosower,
  {\it Planar amplitudes in maximally supersymmetric Yang-Mills theory,}
  Phys.\ Rev.\ Lett.\  {\bf 91} (2003) 251602,
  {\tt hep-th/0309040}.




\bibitem{bds}
  Z.~Bern, L.~J.~Dixon and V.~A.~Smirnov,
  {\it Iteration of planar amplitudes in maximally supersymmetric Yang-Mills
  theory at three loops and beyond,}
  Phys.\ Rev.\  D {\bf 72} (2005) 085001,
  {\tt hep-th/0505205}.







\bibitem{Polyakov:1980ca}
A.~M.~Polyakov,
{\it Gauge Fields As Rings Of Glue},
Nucl.\ Phys.\  B {\bf 164} (1980) 171.

\bibitem{Brandt:1981kf}
  R.~A.~Brandt, F.~Neri and M.~a.~Sato,
  {\it Renormalization Of Loop Functions For All Loops,}
  Phys.\ Rev.\  D {\bf 24} (1981) 879.

\bibitem{Korchemsky:1985xj}
  G.~P.~Korchemsky and A.~V.~Radyushkin,
  {\it Loop Space Formalism And Renormalization Group For The Infrared Asymptotics Of QCD,}
  Phys.\ Lett.\  B {\bf 171} (1986) 459.

\bibitem{bes}
  N.~Beisert, B.~Eden and M.~Staudacher,
  {\it Transcendentality and crossing,}
  J.\ Stat.\ Mech.\  {\bf 0701}, P021 (2007),
  {\tt hep-th/0610251}.


\bibitem{bkk}
  B.~Basso, G.~P.~Korchemsky and J.~Kotanski,
  {\it Cusp anomalous dimension in maximally supersymmetric Yang-Mills theory at
  strong coupling,}
  Phys.\ Rev.\ Lett.\  {\bf 100} (2008) 091601,
  {\tt 0708.3933 [hep-th]}.




\bibitem{am}
  L.~F.~Alday and J.~Maldacena,
  {\it Gluon scattering amplitudes at strong coupling,}
  JHEP {\bf 0706} (2007) 064,
  {\tt 0705.0303 [hep-th]}.




\bibitem{dks}
  J.~M.~Drummond, G.~P.~Korchemsky and E.~Sokatchev,
  {\it Conformal properties of four-gluon planar amplitudes and Wilson loops,}
  Nucl.\ Phys.\  B {\bf 795} (2008) 385,
  {\tt 0707.0243 [hep-th]}.


\bibitem{bht}
  A.~Brandhuber, P.~Heslop and G.~Travaglini,
  {\it MHV Amplitudes in N=4 Super Yang-Mills and Wilson Loops,}
  Nucl.\ Phys.\  B {\bf 794} (2008) 231,
  {\tt 0707.1153 [hep-th]}.



\bibitem{dhks4}
  J.~M.~Drummond, J.~Henn, G.~P.~Korchemsky and E.~Sokatchev,
  {\it On planar gluon amplitudes/Wilson loops duality,}
  Nucl.\ Phys.\  B {\bf 795} (2008) 52,
  {\tt 0709.2368 [hep-th]}.

\bibitem{dhks5}
  J.~M.~Drummond, J.~Henn, G.~P.~Korchemsky and E.~Sokatchev,
{\it Conformal Ward identities for Wilson loops and a test of the duality with
  gluon amplitudes,}
{\tt 0712.1223 [hep-th]}.





\bibitem{Cachazo:2006tj}
  F.~Cachazo, M.~Spradlin and A.~Volovich,
  {\it Iterative structure within the five-particle two-loop amplitude,}
  Phys.\ Rev.\  D {\bf 74} (2006) 045020,
  {\tt hep-th/0602228}.

\bibitem{2l5pt}
  Z.~Bern, M.~Czakon, D.~A.~Kosower, R.~Roiban and V.~A.~Smirnov,
  {\it Two-loop iteration of five-point N = 4 super-Yang-Mills amplitudes,}
  Phys.\ Rev.\ Lett.\  {\bf 97} (2006) 181601,
  {\tt hep-th/0604074}.




\bibitem{Alday:2007he}
  L.~F.~Alday and J.~Maldacena,
 {\it Comments on gluon scattering amplitudes via AdS/CFT,}
  JHEP {\bf 0711} (2007) 068,
  {\tt 0710.1060 [hep-th]}.

\bibitem{dhksbum}
  J.~M.~Drummond, J.~Henn, G.~P.~Korchemsky and E.~Sokatchev,
   {\it The hexagon Wilson loop and the BDS ansatz for the six-gluon amplitude,}
Phys.\ Lett.\  B {\bf 662} (2008) 456,
{\tt 0712.4138 [hep-th]}.


\bibitem{Bartels:2008ce}
  J.~Bartels, L.~N.~Lipatov and A.~S.~Vera,
  {\it BFKL Pomeron, Reggeized gluons and Bern-Dixon-Smirnov amplitudes,}
  {\tt 0802.2065 [hep-th]}.


\bibitem{seven}
  Z.~Bern, L.~J.~Dixon, D.~A.~Kosower, R.~Roiban, M.~Spradlin, C.~Vergu and A.~Volovich,
{\it The Two-Loop Six-Gluon MHV Amplitude in Maximally Supersymmetric Yang-Mills
  Theory,} Phys.\ Rev.\  D {\bf 78}, 045007 (2008),
  {\tt 0803.1465 [hep-th]}.

\bibitem{dhks6}
  J.~M.~Drummond, J.~Henn, G.~P.~Korchemsky and E.~Sokatchev,
  {\it Hexagon Wilson loop = six-gluon MHV amplitude,}
  Nucl.\ Phys.\  B {\bf 815} (2009) 142,
{\tt 0803.1466 [hep-th].}





\bibitem{Anastasiou:2009kn}
  C.~Anastasiou, A.~Brandhuber, P.~Heslop, V.~V.~Khoze, B.~Spence and G.~Travaglini,
 {\it Two-Loop Polygon Wilson Loops in N=4 SYM,}  JHEP {\bf 0905} (2009) 115,
  {\tt 0902.2245 [hep-th]}.




\bibitem{Anastasiou:2008rm}
  C.~Anastasiou, S.~Beerli and A.~Daleo,
  {\it The two-loop QCD amplitude gg $\to$  h,H in the Minimal Supersymmetric Standard
  Model,}
  Phys.\ Rev.\ Lett.\  {\bf 100}, 241806 (2008),
  {\tt 0803.3065 [hep-ph]}.

\bibitem{Anastasiou:2007qb}
  C.~Anastasiou, S.~Beerli and A.~Daleo,
  {\it Evaluating multi-loop Feynman diagrams with infrared and threshold
  singularities numerically,}
  JHEP {\bf 0705}, 071 (2007),
  {\tt hep-ph/0703282}.


\bibitem{Lazopoulos:2007ix}
  A.~Lazopoulos, K.~Melnikov and F.~Petriello,
  {\it QCD corrections to tri-boson production,}
  Phys.\ Rev.\  D {\bf 76}, 014001 (2007),
  {\tt hep-ph/0703273}.

\bibitem{Anastasiou:2005pn}
  C.~Anastasiou, K.~Melnikov and F.~Petriello,
 {\it The electron energy spectrum in muon decay through O(alpha**2),}
  JHEP {\bf 0709}, 014 (2007),
  {\tt hep-ph/0505069}.


\bibitem{cristian-n}
  C.~Vergu,
  {\it The two-loop MHV amplitudes in N=4 supersymmetric Yang-Mills theory,}
 {\tt 0908.2394 [hep-th]}.

\bibitem{Georgiou:2009mp}
  G.~Georgiou,
  {\it Null Wilson loops with a self-crossing and the Wilson loop/amplitude
  conjecture,}
  JHEP {\bf 0909} (2009) 021,
  {\tt 0904.4675 [hep-th]}.


\bibitem{Schabinger:2009bb}
  R.~M.~Schabinger,
 {\it The Imaginary Part of the N = 4 Super-Yang-Mills Two-Loop Six-Point MHV
  Amplitude in Multi-Regge Kinematics,}
  {\tt 0910.3933 [hep-th]}.


\bibitem{amsmalloctopus}
  L.~F.~Alday and J.~Maldacena,
  {\it Minimal surfaces in AdS and the eight-gluon scattering amplitude at strong
  coupling,}
  {\tt 0903.4707 [hep-th].}



\bibitem{amoctopus}
  L.~F.~Alday and J.~Maldacena,
 {\it Null polygonal Wilson loops and minimal surfaces in Anti-de-Sitter space,}
  {\tt 0904.0663 [hep-th]}.



\bibitem{Gaiotto:2008cd}
  D.~Gaiotto, G.~W.~Moore and A.~Neitzke,
  {\it Four-dimensional wall-crossing via three-dimensional field theory,}
  {\tt 0807.4723 [hep-th]}.




\bibitem{bddk}
Z.~Bern, L.~J.~Dixon, D.~C.~Dunbar and D.~A.~Kosower, {\it One
Loop N Point Gauge Theory Amplitudes, Unitarity And Collinear
Limits,} Nucl.\ Phys.\ B {\bf 425} (1994) 217, {\tt
hep-ph/9403226}.


  \bibitem{fusing}
  Z.~Bern, L.~J.~Dixon, D.~C.~Dunbar and D.~A.~Kosower,
 {\it Fusing gauge theory tree amplitudes into loop amplitudes,}
  Nucl.\ Phys.\  B {\bf 435}, 59 (1995),
  {\tt hep-ph/9409265}.



\bibitem{kk}
  I.~A.~Korchemskaya and G.~P.~Korchemsky,
  {\it On lightlike Wilson loops,}
  Phys.\ Lett.\  B {\bf 287} (1992) 169.















\end{thebibliography}
\end{document}